\documentclass[a4paper, 12pt]{article}
\usepackage[latin1]{inputenc}      
\usepackage[top=2.5cm, bottom=2cm, left=2.5cm, right=3cm]{geometry}
\usepackage{graphicx}                  
\usepackage{float}
\usepackage{amsfonts}
\usepackage{lscape}
\usepackage{epstopdf}
\usepackage{pdflscape}
\usepackage{amsbsy} 
\usepackage{amsmath,amssymb}
\usepackage{array,booktabs,calc}
\usepackage[mathscr]{euscript}
\usepackage{natbib}
\usepackage[colorlinks=true,linkcolor=blue, citecolor=blue]{hyperref}%

\usepackage{color}
\usepackage[toc,page]{appendix}
\usepackage{booktabs}
\usepackage{pdfpages}
\usepackage{tikz}
\usetikzlibrary{automata, positioning}
\usepackage{multirow}
\usepackage{lscape}

\begin{document}

\title{\textbf{Bayesian cross-validation of geostatistical models}}
\date{}
\author{Viviana G. R. Lobo$^{1,2}$\footnote{\texttt{viviana@dme.ufrj.br}}         \and
        Thaís C. O. Fonseca${^1}$\footnote{\texttt{thais@im.ufrj.br},  Av. Athos da Silveira Ramos, Centro de Tecnologia, 
             Bloco C Sala C114 D, IM-UFRJ, CEP 21941-909, 
             Rio de Janeiro,
             Brazil} \and
        Fernando A. S. Moura$^{1}$\footnote{\texttt{fmoura@im.ufrj.br}} \and  ${}^{1}$ Federal University of Rio de Janeiro, Rio de Janeiro, Brazil \\
   ${}^{2}$ Federal University Fluminense, Niterói, Brazil}

\maketitle

\begin{abstract}
The problem of validating or criticising models for georeferenced data is challenging, since the conclusions can vary significantly depending on the locations of the validation set. This work proposes the use of cross-validation techniques to assess the goodness of fit of spatial models in different regions of the spatial domain to account for uncertainty in the choice of the validation sets. An obvious problem with the basic cross-validation scheme is that it is based on selecting only a few out of sample locations to validate the model, possibily making the conclusions sensitive to which partition of the data into training and validation cases is utilized. A possible solution to this issue would be to consider all possible configurations of data divided into training and validation observations. From a Bayesian point of view, this could be computationally demanding, as estimation of parameters usually requires Monte Carlo Markov Chain methods. To deal with this problem, we propose the use of estimated discrepancy functions considering all configurations of data partition in a computationally efficient manner based on sampling importance resampling. In particular, we consider uncertainty in the locations by assigning a prior distribution to them. Furthermore, we propose a stratified cross-validation scheme to take into account spatial heterogeneity, reducing the total variance of estimated predictive discrepancy measures considered for model assessment. We illustrate the advantages of our proposal with simulated examples of homogeneous and inhomogeneous spatial processes to investigate the effects of our proposal in scenarios of preferential sampling designs. The methods are illustrated with an application to a rainfall dataset.\\

\end{abstract}


\section{Introduction}\label{intro}

In many practical problems, the researcher is interested in modelling some phenomenon that occurred in space as a stochastic process. Goodness of fit of the assumed model is an important step. However, in the geostatistical context, this is a challenging problem since only one realization of the process is available for both parameter estimation and model checking. \\

\cite{Diggle2014} points out that if a model fits  the data well, it can be used to generate datasets which are statistically similar to the observed sample. In addition, the practioner may be interested in making predictions for out-of-sample locations from the fitted model. These practical statistical problems indicate that cross-validation techniques are  potentially useful tools for model checking. \\

The usual approaches for model checking in spatial statistics are based on selecting a subset from the locations to make prediction with the assumed model. The observed values which were left out of the estimation procedure are then compared with the predicted values. However, the choice of locations used for model fitting and prediction is usually ad-hoc. We are interested in assessing the goodness of fit of spatial models using cross-validation tools, as well as allowing for model checking in different regions of the spatial domain. \\

From a theoretical point of view, statistical inference goes beyond parameter estimation and prediction \citep[see][page 343]{Rob07}. Often, tests are performed regarding model parameters that are based on models that are not adequate to the data under study. That is, model adequacy checking should not be based on model parameter testing. Some verification of model goodness of fit is then called for. From a Bayesian perspective, statements are made regarding the posterior distribution, which are also based on the chosen sampling distribution of the data. The usual model criticism is done through model comparison and prediction for a few out-of-sample observations. Often, these model checks are not able to assess whether the assumed model is plausible for the data in the whole spatial domain. \\

In the literature, various authors have suggested the use of cross-validation for modelling univariate data. \cite{Burman89} introduces validation techniques in a study of optimal transformation of variables, based on ${k}$-fold cross-validation and repeated learning testing methods. \cite{Thall1997} demonstrate that repeated data splitting is preferred over ${k}$-fold cross-validation. They propose to apply cross-validation to a very large number of randomly generated partitions of the data. \\

From a Bayesian standpoint, \cite{Marshall2003} and \cite{Burman89}, amongst others, show that the cross-validation can be computationally very expensive, since a full MCMC analysis has to be repeated, leaving out each in turn validation set. \cite{Cressie2000} consider importance weighting and resampling methods in the context of posterior predictive model checking via conditional predictive ordinate (CPO) and posterior predictive p value. In the univariate setting, \cite{alqgust2001} propose Bayesian cross-validation for several data partitions sampled from the prior distribution of the possible partition into training and validation sets. The model checking is based on estimating discrepancy functions, which are statistical measures commonly used in the literature for model comparison. Although many papers have exploited cross-validation methods for univariate data, this is not the case for spatial data analysis. For instance, the usual setup for model checking in geostatistics is to make prediction for one or a few selected validation sets. \cite{Cressie93} states that the basic cross-validation idea in the spatial setup is to delete some of the data and use the remaining data to predict the deleted observations. Then the prediction error can be inferred from the predicted and observed values. If the model adequately described the implicit spatial dependence on the dataset, then the predicted value should be close to the true value.\\

However, the choice of observation sites in a spatial process is not always ``robust'' to the considered sampling or allocation of sites. In general, it does not consider the sampling process that generated the locations. In fact, models that ignore information about sample selection can lead wrong inferences and, therefore, to wrong predictions. In that context, we propose to allow for uncertainty in the selection of the validation sets by considering a prior distribution for the spatial locations. In particular, the prior assigned is uniform on the spatial domain of interest. This process is considered to be homogeneous or homogeneous in sub-regions to account for possible preferential sampling in the locations.\\

The use of cross-validation techniques in a large volume of spatial data becomes a computational challenge, due to the difficulty of applying traditional prediction methods in a time-tolerant boundary. If we were to make prediction for several vectors of points, the cross-validation procedure would be repeated again for all possible selected configurations of training and validation samples. For most geostatistics problems, this scheme becomes computationally prohibitive. Thus, more sophisticated approach are useful, both to reduce the final cost and increase efficiency. To deal with this problem, we propose an efficient algorithm for cross-validation, which is based on importance weighting and only a handful of MCMC runs.\\

This paper is organized as follows. In Section 2, the main motivation of this work is presented. Section 3 briefly reviews the main aspects of spatial data analysis. Section 4 describes the Bayesian cross-validation of the spatial data using estimators of expected discrepancy functions via MCMC (\textit{Monte Carlo Markov Chain}). Given this foundation, Section 5 reports a procedure for validating models based on stratified spatial data. In Section 6, simulated examples are presented. Finally, Section 7 shows an application to a rainfall dataset. \\

\section{Motivation}\label{sec2}

\textbf{\subsection{Uncertainty of the data partition}}

In general in geostatistical inference, to evaluate an assumed model some locations are removed from the dataset, the so-called validation set, and then observed data and predicted values are compared through some discrepancy function. The choice of the locations to be removed from the data for validation purposes depends on the spatial arrangement and is usually done in a random manner. In this work, the spatial locations are assumed to be a random sample from a specified distribution and locations are sampled to compose the validation set according to the prior probabilities of the sets.\\

The predictive performance of a model is typically based on the assumption that the reference model is believed to best describe the knownledge about future observations. In the scope of geostatistical modelling, the basic component are spatial locations and data observed at these locations. According to \cite{Cressie93}, the data, or even sometimes the locations, are assumed to be random. In this circumstance, the use of adequate spatial models is crucial for prediction.\\

As discussed in \cite{Berger2004}, the important objective of approaches to Bayesian model selection is the use of training samples. However, the choice of the training data is not trivial. In particular, this is most relevant in the spatial data analysis context, because different sets of training data might lead to very different inferences.\\

Our main contribution is to develop an efficient computing procedure to make cross-validation of statistical models for spatially correlated data. Thus, we extend the work of \cite{alqgust2001} to correlated data modelling, in particular, the spatial dependence and training and validation sets uncertainty are take into account by allowing the locations to be realizations of a spatial point process.\\

The way the phenomenon is observed may be related to the intensity of the process. In this context, we consider the spatial heterogeneity as being a feature of a natural phenomenon. The heterogeneity comprises two elements of second-order variation: global variance and the spatial autocorrelation. We adopt spatial stratified sampling, where that are the heterogeneous area is divided into several subareas more homogeneous than the whole area, reducing the total variance of estimated predictive discrepancy measures considered for model assessment.

\textbf{\subsection{An illustrative example}}

Let us consider an illustrative example simulated in a unit square, with 52 data points randomly located within the area (irregular grid). The simulated values were generated according to a Student-t process with $\nu=3, \sigma^2=1$, $\mu=2$ and $\phi=0.3$. Therefore, $Y= \left( Y(x_1), \ldots, Y(x_n) \right)$ is distributed as
\vspace{-0.05cm}
\begin{equation}
\label{eq:eq12}
 Y \mid \mu,\sigma^2,\phi, \nu \sim ST \left(\mu\mathbf{1}, \nu, \sigma^2 R \right)  
\end{equation}

\noindent where $R$ is an exponential correlation matrix, with range parameter $\phi$. For prediction purpose, we choose 100 randomly selected validation configurations of all possible subsets of size 3. Notice that there are $\binom{52}{3}$ possible validation sets, thus performing cross-validation for all possibilities is too time consuming. For this arbitrarily chosen configurations], we omitted three points (validation sampling) and calculated the predicted value for these locations using the remaining points (training sampling). We fitted the Gaussian and Student-t spatial models to the simulated data and used MCMC techniques for estimating the model parameters. The prior distribution assigned to $\mu$, $\sigma^2 $, $\phi$ and $\nu$, can be seen in Appendix 1.\\

For model assessment, we considered the mean square prediction error as the discrepancy measure, which is given by $\frac{\sum_{i=1}^{n}(\hat{y}_{i} - {y}_{i})^2}{n}$.
\begin{figure}[H]
\begin{center}
\begin{tabular}{cc}
\includegraphics[width=0.25\textwidth]{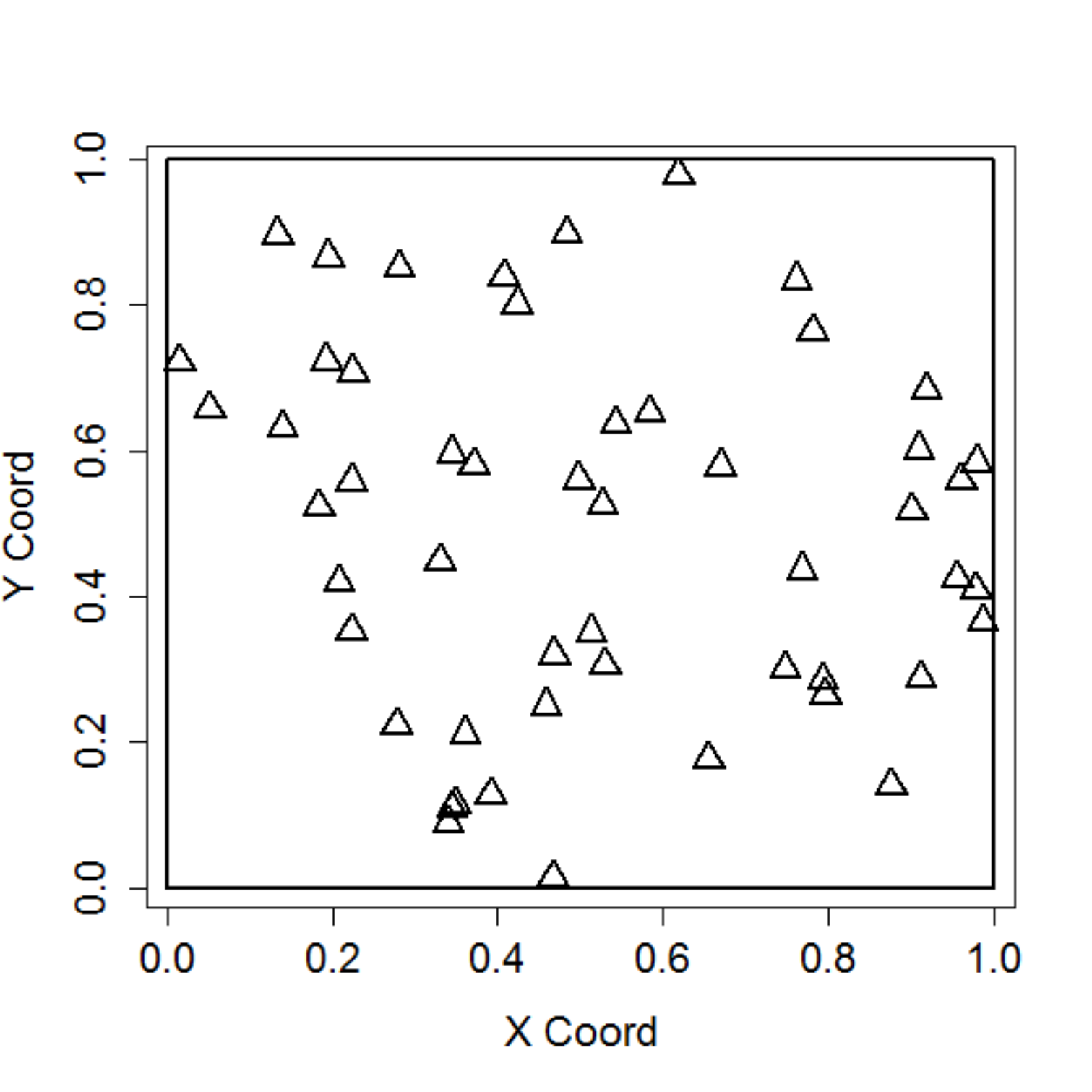} 
&
\includegraphics[width=0.25\textwidth]{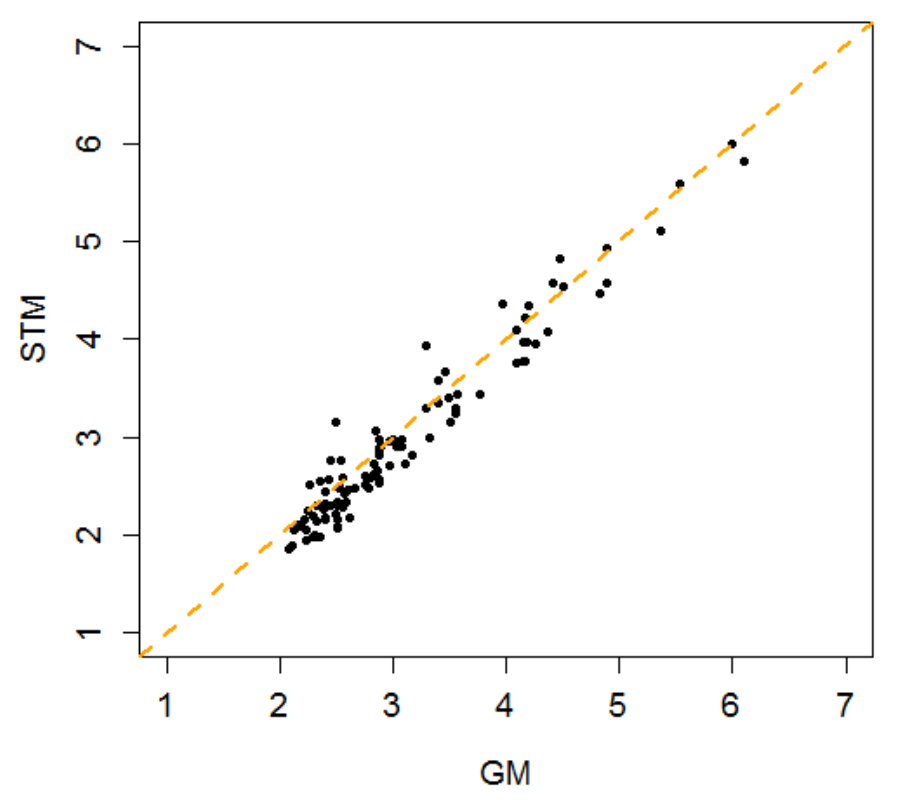} 
\\
(a) sample locations &  (b) predictive discrepancy measure\\
\end{tabular}
\caption{ (a) sample locations generate randomly for given covariance ($\sigma^2=1.0, \phi=0.3$) parameters and $\mu=2, \tau^2=0$, $\nu=3$. (b) cross-validation performance: Gaussian model (GM) versus Student t model (STM) for each validation configuration (100) using mean square prediction errors.}
\label{fig:sec1fig3}
\end{center}
\end{figure}

\begin{figure}[H]
\begin{center}
\begin{tabular}{ccc}
\includegraphics[width=0.25\textwidth]{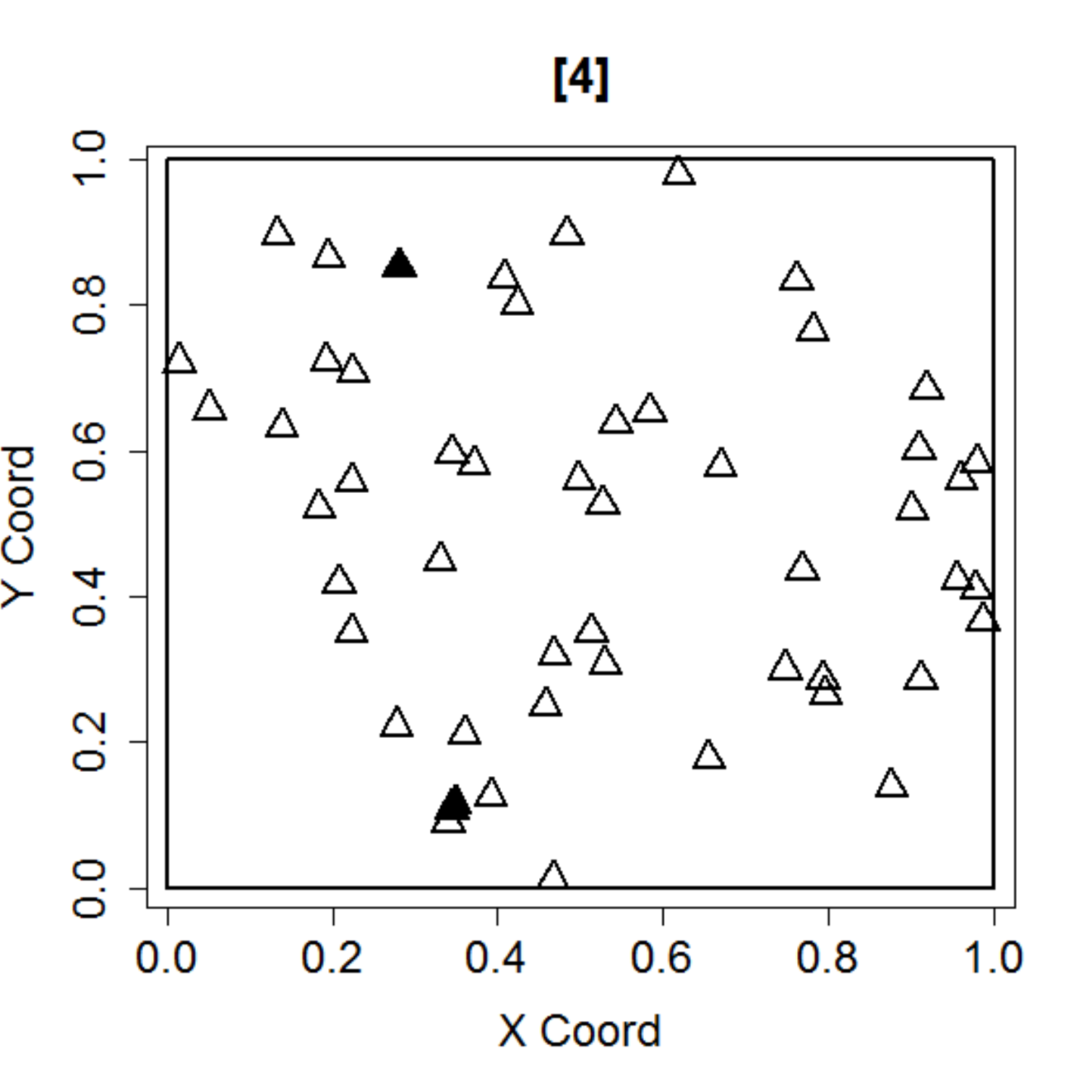} 
&
\includegraphics[width=0.25\textwidth]{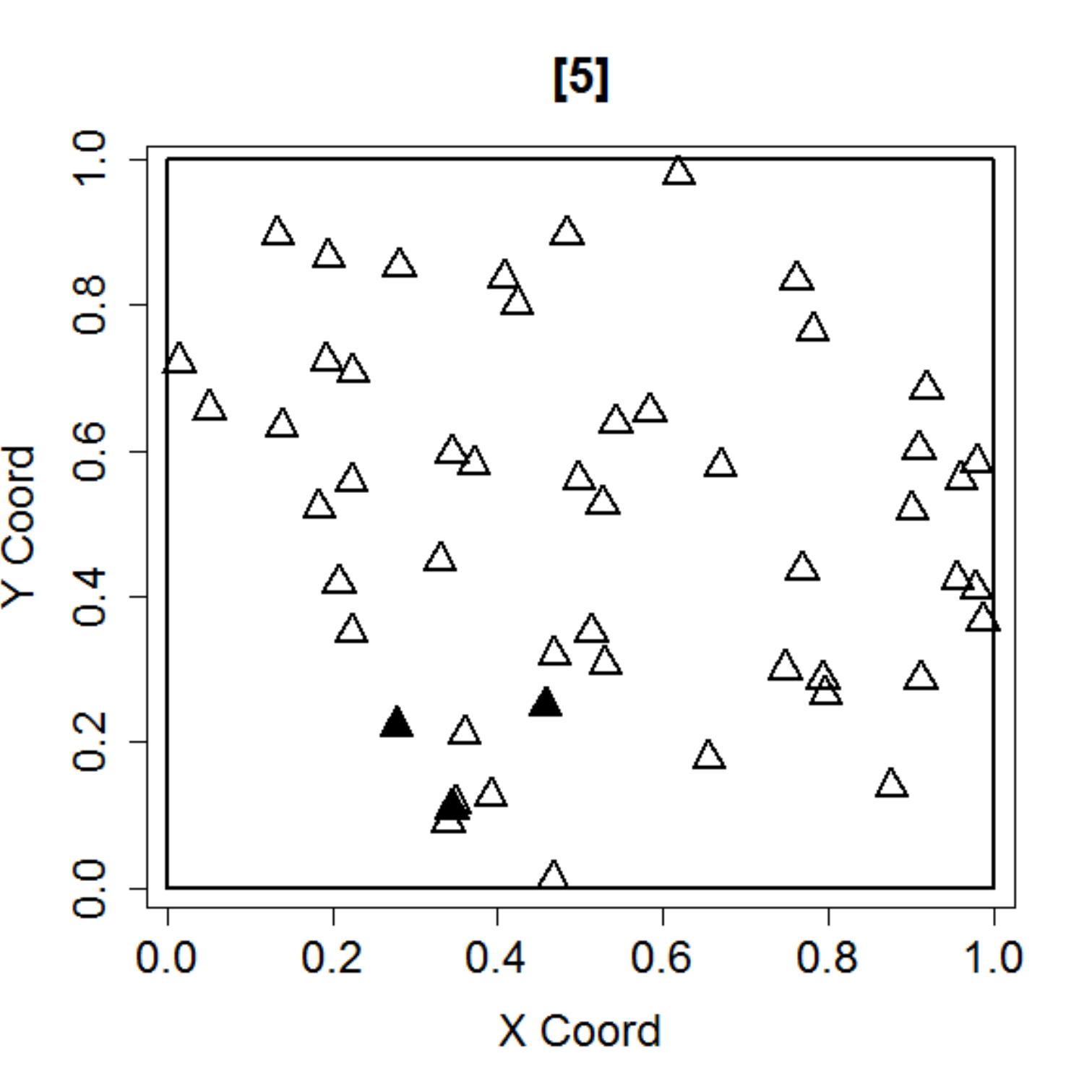}
&
\includegraphics[width=0.25\textwidth]{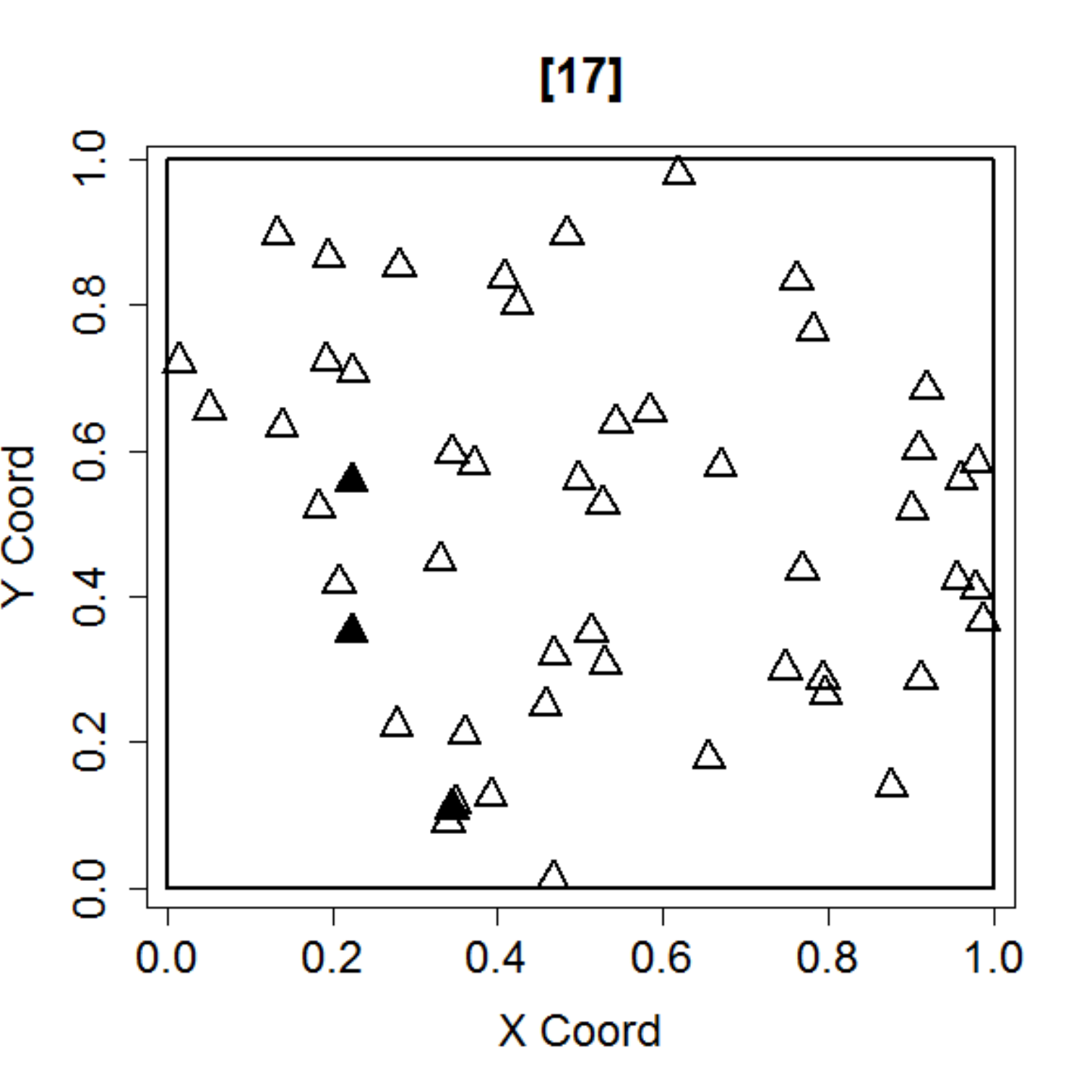}
\\
\includegraphics[width=0.25\textwidth]{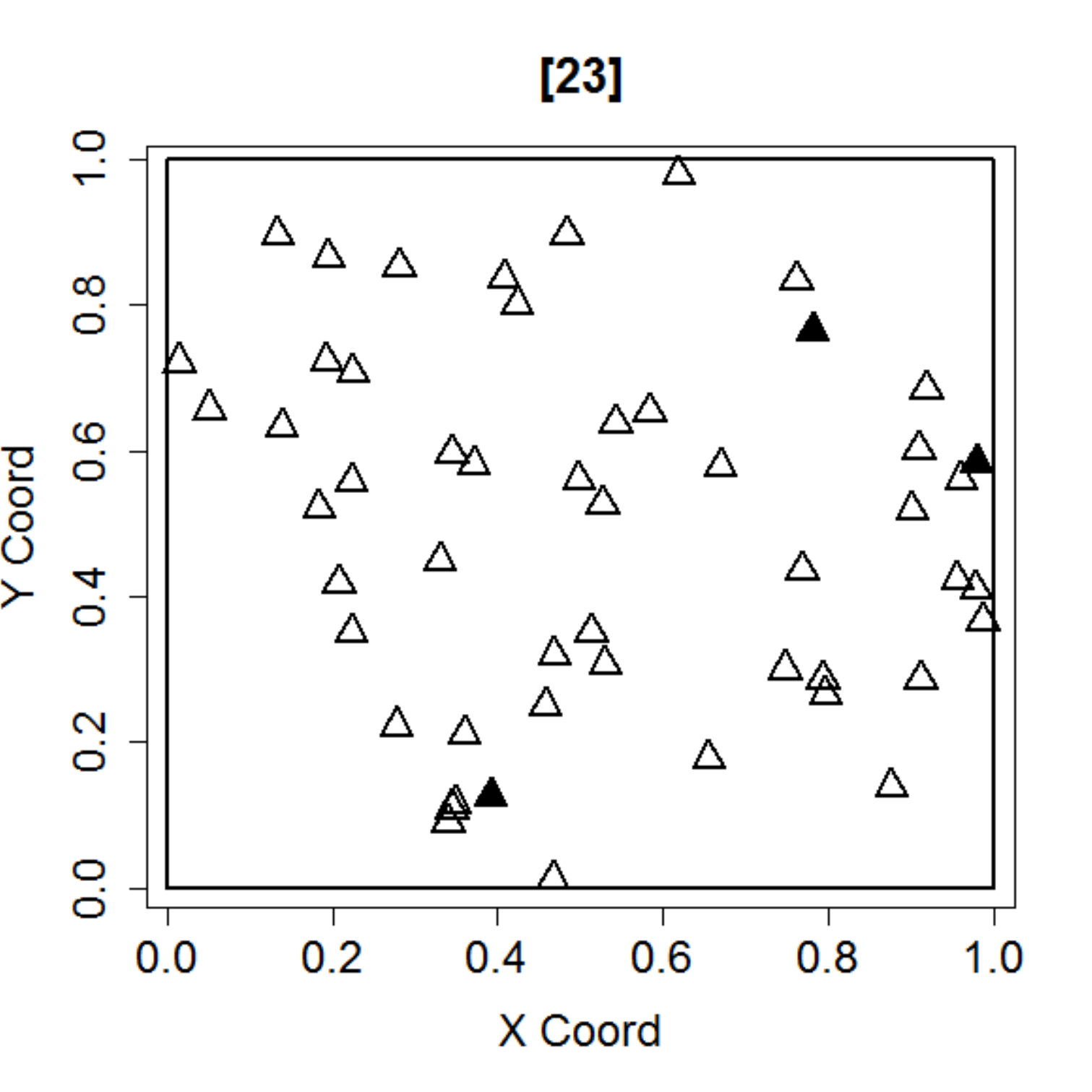} 
&
\includegraphics[width=0.25\textwidth]{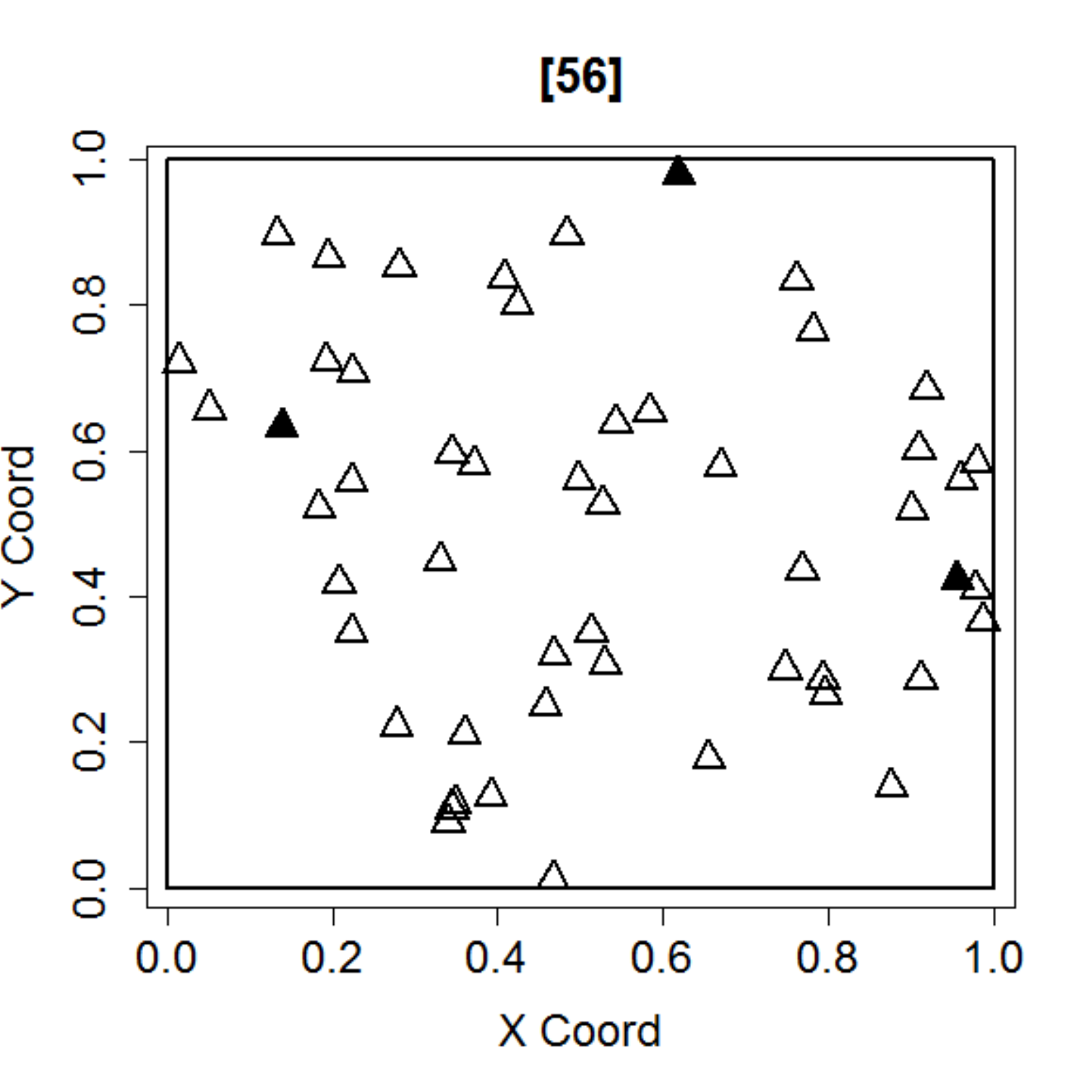}
&
\includegraphics[width=0.25\textwidth]{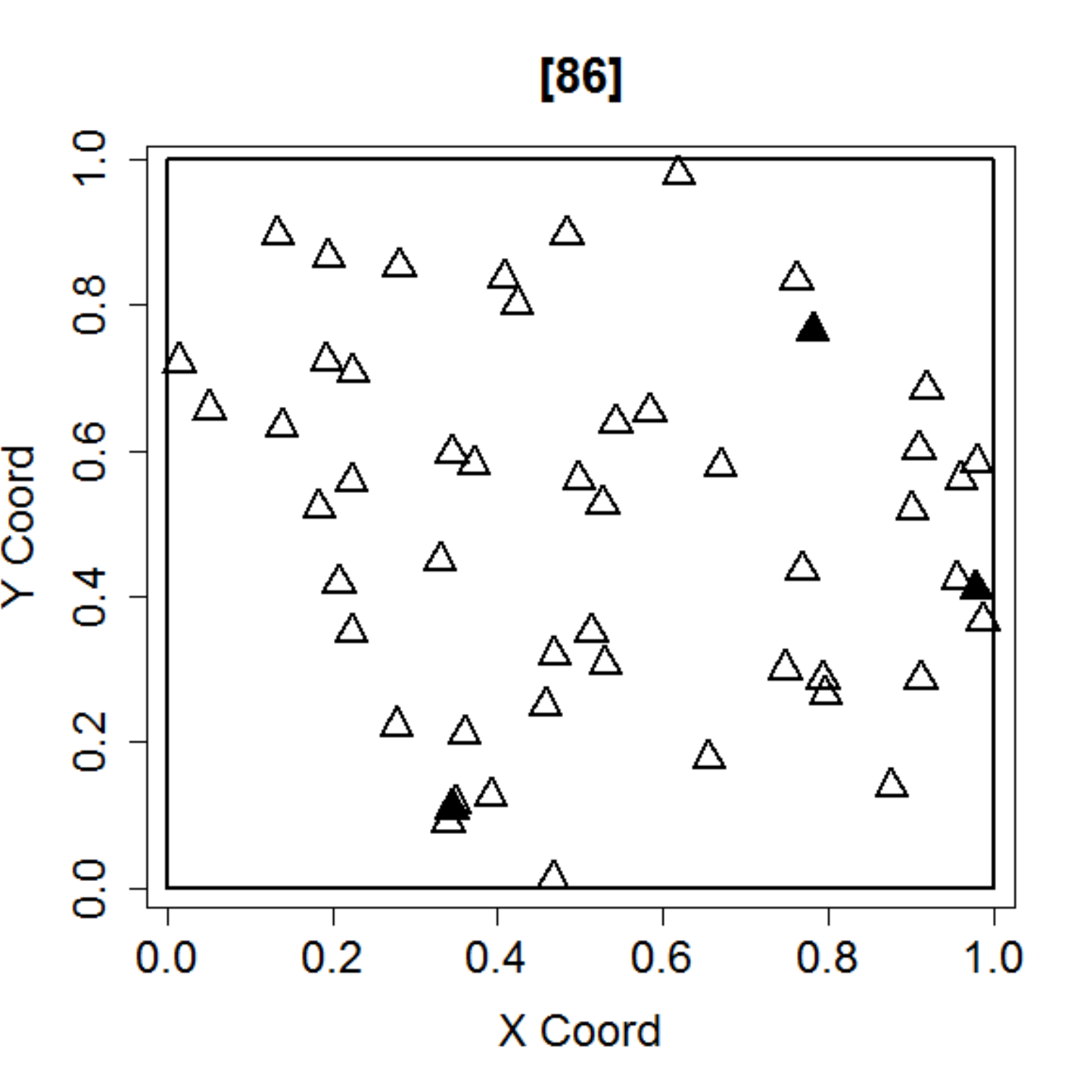}
\end{tabular}
\caption{Some configurations of training locations (empty triangles) and validation locations (black triangles), where the Gaussian model was chosen as the best model for the Student-t model illustrative dataset.}
\label{fig:sec1fig3a}
\end{center}
\end{figure}

Figure \ref{fig:sec1fig3} presents 100 randomly selected validation configurations for cross-validation performance. Notice that we have different results for each choice of validation sample. Although the process was generated from the Student-\textit{t} distribution, the performance of the Student-t model is worse in 26\% of the validation sets when compared to the Gaussian model. This example motivates our work to incorporate the uncertainty in the validation set.\\

Figure \ref{fig:sec1fig3a} presents some configurations of training locations and validation locations where the Gaussian model was chosen as the best model for the Student-t illustrative dataset. Notice that most of these validation sets have sites on the border of the spatial region.\\

According to \cite{Stan1997}, the similar inferences are made about the mean $\mu$ under Gaussian model (GM) and Student-t model (STM), but different inferences can be made about scale, because the scale is differently represented in the two models. Thus, each model used for prediction might lead to different prediction intervals for ungauged locations.  As discussed in \cite{alqgust2001}, an obvious problem with the basic cross-validation scheme is that the results may be sensitive to which particular partition of the data into training and validation cases is utilized. In some circumstances, predictive assessments may even be sensitive to how the data are segmented for $k$-fold cross-validation (\cite{Burman89}).\\

From this illustrative example, we propose an approach that is able to verify the goodness of fit of a model for different configurations of data validation sampling and still preserve computational feasibility.

\section{Basic geostatistical model}\label{SecModGeo}

In this section we present the essential elements of spatial models for geostatistical data analysis. \\

The fundamental concept underlying the theory according to \cite{ban2004Hier} is a stochastic process $\left\{ Y(x): x \in A \right\}$, where $A$ is a fixed subset of a $d$-dimensional Euclidean space. In the spatial context, we usually have $d$ equal to 2 (northings and eastings) or 3 (northings, eastings and altitude). We consider situations where $d>1$ and refer to them as \textit{a spatial process}.\\

Let us consider a finite set of spatial sample locations ${x}= (x_1, x_2, \ldots, x_n)$ within a region $A$. Geostatistical data consist of measurements taken at the sample locations ${x}$. Therefore, we denote the data vector by $(y(x_1), y(x_2), \ldots, y(x_n))$, the observed values of a random vector $Y$. We assume that ${x}$ varies continously throughout the region $A$.\\

According to \cite{Diggle2010}, the data are obtained by sampling a spatially continuous phenomenon $S(x)= x \in \mathbb{R}^2$ at a discrete set of locations $x_i=1, \ldots, n$ in a spatial region of interest $A \subset \mathbb{R}^2$. Hence, if $Y_i$ denotes the measured value at the location $x_i$, a simple model for the data takes the form 

\begin{equation}\label{eq:sec2eq1}
Y_i = \mu + S(x_i) + Z_i \quad i=1, \ldots,n,
\end{equation}

\noindent where $\mu$ represent the mean and the $Z_i's$ are mutually independent, zero-mean random variables with variance $\tau^2$. The underlying spatial process $\left\{ S(x): x \in \mathbb{R}^2 \right\}$ is a stationary Gaussian process with zero mean, constant variance $\sigma^2$ and correlation function $\rho({u}, \phi)$, where $\phi$ is the correlation function parameter and ${u}$ is the distance between two locations.


\section{Cross-validation of Bayesian models for spatially correlated data}\label{Sec4}

We consider the uncertainty in the choice of data split into validation and training sets by defining a prior distribution of such sets. In the Bayesian analysis of spatial data, the fit of the model usually requires MCMC sampling from the posterior distribution. We extend the technique proposed by \cite{alqgust2001} to spatially correlated data, so the validation measure does not require a separate posterior sample for each training sample.

\textbf{\subsection{The reference distribution for spatial data }}

Suppose that the observed data consist of responses $\mathbf{y}= (y_1, y_2, \ldots, y_n)$ arising from the process $Y(x)$  and locations $X = (x_1,  \ldots, n_n)$ described in equation \eqref{eq:sec2eq1}. Our cross-validation assessments are based on expectations under a \textit{reference distribution} for $(\mathbf{s}_{y}, \mathbf{y}, \theta, \mathbf{y}^{rep})$, following \cite{alqgust2001}, where the \textit{split} $\mathbf{s}_{y}$ is a 0 -- 1 vector which divides the $n$ cases into training and validation cases. The \textit{parameter} $\theta$ describes the statistical model and the \textit{replicated response} $\mathbf{y}^{rep}$ is a hypothetical realization of the response vector. \\

The reference distribution is most easily defined via factorization

\begin{equation}
\label{eq:eqreference}
p(\mathbf{s}_{y},\mathbf{y},\theta, \mathbf{y}^{rep}) = p(\mathbf{s}_{y}) p( \theta \mid \mathbf{s}_{y},\mathbf{y}) p(\mathbf{y}^{rep} \mid \theta, \mathbf{s}_{y},\mathbf{y}).
\end{equation}

 We consider the prior $p(\mathbf{s}_{y})$ to be the uniform distribution over such splits. Notice that this choice of prior might not be reasonable if there is a pattern in the $X= (x_1, \ldots, x_n)$ generation, as in the case of inhomogeneous processes. We consider an extension of this prior in Section \ref{sec4.5}.\\

For the purpose of building the split vectors, we denote the sample sizes of training ($0$ - \textit{zero}) and validation ($1$ - \textit{one}) by $n_T$ and $n_V$, respectively. In spatially correlated data, we define the specific split vector as 

$$s_{l}= \left \{
\begin{array}{ll}
     0, &  x_l \mbox{ is a training location}\\
     1,              & \mbox{otherwise},
\end{array}
\right .$$

\noindent and split vector $\mathbf{s}_{y}= (s_1, \ldots, s_n)$ are vectors of the same dimension of $\mathbf{y}$, indicating for all locations $x_1, \ldots, x_n$ where $y_l$, $l= 1, \ldots, n$ is used for training or validation.\\

After the choice of a specific split vector $\mathbf{s}_y$, let $\mathbf{y}_{T[\mathbf{s}_y]}$ and $\mathbf{y}_{V[\mathbf{s}_y]}$ be defined as the observed training and validation cases, with length $n_T$ and $n_V$ respectively. Given the split $\mathbf{s}_{y}$, $p(\theta \mid \mathbf{s}_{y}, \mathbf{y})$ is defined as the posterior distribution of $\theta$ given the training data only. Thus, using Bayes theorem: 

\begin{equation}\label{eq:sec2eq3}
p(\theta \mid \mathbf{s}_{y}, \mathbf{y}) \propto f(\mathbf{y}_{T[\mathbf{s}_y]} \mid  \theta) \pi(\theta).
\end{equation}

Notice that $\mathbf{y}^{rep}$ is simply distributed according to the sampling model assumed for the data, i.e., $[\mathbf{y}^{rep} \mid \theta, \mathbf{s}_{y}, \mathbf{y}]$. Thus, under the reference distribution, this is $[\mathbf{y}_{V[\mathbf{s}_{y}]} \mid \mathbf{y}_{T[\mathbf{s}_{y}]}, \theta]$, which represents the predictive distribution of the validation given the training data, for a specific split vector $\mathbf{s}_{y}$.

\textbf{\subsection{Expected discrepancy estimation}\label{efficient}}

The cross-validation assessments are based on the expected discrepancy to be computed under a reference distribution. It requires the distribution of the replicated response vector $\mathbf{y}^{rep}$. In particular, we are interested in computing expectations, generically denoted by
\begin{equation}\label{eq:eq11}
{\Psi} = E\left\{ r(\mathbf{s}_{y}, \mathbf{y}, \theta, \mathbf{y}^{rep})\right\}.
\end{equation}

\noindent The expected value in \eqref{eq:eq11} represents a statistical measure for comparing Bayesian models and $r$ represents a discrepancy function. The expected value can be calculated as, 

\small
\begin{eqnarray} \nonumber
{\Psi} &=&  \int_{\Theta}\int_{Y}\sum_{S} r(\mathbf{s}_{y}, \mathbf{y}, \theta, \mathbf{y}^{rep}) p(\mathbf{s}_{y}, \mathbf{y}, \theta, \mathbf{y}^{rep}) d\theta d \mathbf{y}^{rep} \\ \nonumber
       &=& \int_{\Theta}\int_{Y}\sum_{S} r(\mathbf{s}_{y}, \mathbf{y}, \theta, \mathbf{y}^{rep}) p(\mathbf{y}^{rep}, \theta \mid  \mathbf{s}_{y}, \mathbf{y} )p(\mathbf{s}_{y})  d\theta d\mathbf{y}^{rep} \\ \nonumber
       &=& \sum_{S} p(\mathbf{s}_{y}) \int_{\Theta} \int_{Y}   r(\mathbf{s}_{y}, \mathbf{y}, \theta, \mathbf{y}^{rep})  p(\mathbf{y}^{rep}, \theta \mid \mathbf{s}_{y}, \mathbf{y}  ) d\theta d \mathbf{y}^{rep} \\ \nonumber
       &=& \sum_{S} E\left\{ r(\mathbf{s}_{y}, \mathbf{y}, \theta, \mathbf{y}^{rep}) \thinspace \vert \thinspace \mathbf{s}_{y} \right\} \thinspace p(\mathbf{s}_{y}) \\ \nonumber
       &  =  & E\left\{ E \left[ r(\mathbf{s}_{y}, \mathbf{y}, \theta, \mathbf{y}^{rep}) \thinspace  \vert \thinspace  \mathbf{s}_{y} \right]  \right\}. \\ \nonumber
\end{eqnarray}

\normalsize
The Monte Carlo approximation to this expected discrepancy under the uniform prior for the splits $s_{\mathbf{y}}$ is given by

\begin{equation}\label{gold}
{\Psi} \approx \frac{1}{I} \sum_{i=1}^{I} E \left\{r(\mathbf{s}_{y}, \mathbf{y}, \theta, \mathbf{y}^{rep}) \mid  \mathbf{s}_{y} = \mathbf{s}_{y}^{(i)}  \right\}.
\end{equation}

The split vectors $\mathbf{s}_{y}^{(1)}, \mathbf{s}_{y}^{(2)} \ldots,\mathbf{s}_{y}^{(I)}$ are simulated independently from $p(\mathbf{s}_{y})$ and $I$ represents the number of splits. Algorithm 1 describes how to compute \eqref{gold} by simulation using Monte Carlo and MCMC simulations from the posterior distribution of model parameters. This estimation is based on obtaining one MCMC sample for each split. We call this estimator the \textit{MC estimator} and it is given by 

\begin{equation}
\label{eqS}
\hat{\Psi}_{mc} = \frac{1}{I}\sum_{i=1}^{I} \frac{1}{J}\sum_{j=1}^{J} r(\mathbf{s}_{y}^{(i)}, \mathbf{y
}, \theta_{ij}, y_{ij}^{rep}),
\end{equation}

\noindent which is an unbiased estimator of expression \eqref{eq:eq11}. \\

Notice that when $p(\theta \mid \mathbf{s}_{y}, \mathbf{y})$ does not have a closed form, MCMC methods can be used to simulate a sample from $p(\theta \mid  \mathbf{s}_{y}, \mathbf{y})$. Next, we present some alternatives to \eqref{eq:eq11}, which are computed from samples from the posterior $p(\theta \mid \mathbf{y})$ and $p(\mathbf{y}^{rep} \mid \theta)$. The number of splits $I$ and the size of the posterior sample for each split $J$ must be specified. The estimator in \eqref{eqS} takes too much time to compute, since each sampled split $\mathbf{s}_{y}$ requires many MCMC runs to sample from $p(\theta \mid \mathbf{s}_{y}, \mathbf{y})$.\\

\begin{flushleft}
\begin{center}
\begin{tabular}{l}
\hline
\small \textbf{Algorithm 1} MC estimator \\
\hline
\small 1: Simulate independent split vectors $\mathbf{s}_{y}^{(1)},\mathbf{s}_{y}^{(2)}, \ldots, \mathbf{s}_{y}^{(I)}$ \\
\small from $p( \mathbf{s}_{y})$.\\
\small 2: For each $\mathbf{s}_{y}^{(i)}$, use a MCMC run to draw a sample \\
\small $\theta_{i1}, \ldots, \theta_{iJ}$ from $p(\theta \mid \mathbf{s}_{y}=\mathbf{s}_{y}^{(i)}, \mathbf{y})$,\\
\small 3: For each $(i,j)$, simulate $y_{ij}^{rep}$ from $p(\mathbf{y}^{rep} \mid \theta= \theta_{ij},\mathbf{s}_{y}, \mathbf{y})$\\
\hline
\end{tabular}
\end{center}
\end{flushleft}

Aiming to reducing the computational cost, we consider the \textit{importance sample estimator}, which requires only a handful of MCMC runs as an alternative estimate of expression \eqref{eq:eq11}. The idea is to approximate the posterior density of a given training sample by a distribution that is based heuristically on the same amount of data, but which does not depend on the specific split $\mathbf{s}_{y}$. In particular, this distribution is  used as an importance function and is defined as

\begin{equation}\label{eq:sec3eq14}
g(\theta) \propto f(\mathbf{y} \mid \theta)^{\alpha} \pi(\theta),
\end{equation}

\noindent where $f(\mathbf{y}\mid \theta) $ denotes the likelihood function for the complete data, $\pi(\theta)$ is the prior distribution and $\alpha= n_T/n$. Sampling importance resampling is considered to obtain a sample from the desired posterior distribution from the approximation \eqref{eq:sec3eq14}. 

\begin{flushleft}
\begin{center}
\begin{tabular}{l}
\hline
\small \textbf{Algorithm 2} SIR estimator \\
\hline
\small 1: Simulate independent split vectors $\mathbf{s}_{y}^{(1)},\mathbf{s}_{y}^{(2)}, \ldots, \mathbf{s}_{y}^{(I)}$ \\
\small from $p( \mathbf{s}_{y})$.\\
\small 2: Let $\theta_{h1}, \ldots, \theta_{hJ}$ be the \textit{h}th of $H$ independent \\
\small  MCMC samples simulated from $g(\theta)$ \\
\small 3: Draw $y_{hj}^{rep}$ from $p(\mathbf{y}^{rep} \mid \theta= \theta_{hj}, \mathbf{y})$, for $h=1, \ldots, H$\\
\small  and $j=1, \ldots, J$ \\
\small 4: Each of these $H$ samples yields an importance sampling\\
\small estimate of $E[r(\mathbf{s}_{y}, \mathbf{y}, \theta, \mathbf{y}^{rep}) \mid  \mathbf{s}_{y}= \mathbf{s}_{y}^{(i)}]$. \\
\hline
\end{tabular}
\end{center}
\end{flushleft}

The SIR estimator is defined as the average of importance sampling estimate of $E[r(\mathbf{s}_{y}, \mathbf{y}, \theta, \mathbf{y}^{rep}) \mid  \mathbf{s}_{y}= \mathbf{s}_{y}^{(i)}]$ across the $I$ independent splits and the $H$ independent samples from $g(\theta)$,

\begin{equation}
\label{eqB}
\hat{\Psi}_{sir} = \frac{1}{H} \sum_{h=1}^{H} \frac{1}{I} \sum_{i=1}^{I} \frac{\sum_{j=1}^{J} r(\mathbf{s}_{y}^{(i)}, \mathbf{y}, \theta_{hj}, y_{hj}^{rep}) w_{hj}}{\sum_{j=1}^{J} w_{hj}}
\end{equation}

\noindent where each weight term $w_{hj}= p(\theta_{hj} \mid \mathbf{s}_{y}^{(i)}, \mathbf{y})/g(\theta_{hj})$ has simple form\footnote{The weights are calculated in Appendix 2}

$$log (w_{hj})=  log f(\mathbf{y}_{T[s]} \mid \theta_{hj}) - \alpha \thinspace  log f(\mathbf{y} \mid \theta_{hj}).$$ 

The number of splits $I$, the size of the posterior sample for each split $J$ and the $H$ independent MCMC samples must be specified.\\

Note that if the simulation standard error is not required, then in fact this estimator can be based on a single MCMC run, i.e., $H=1$, otherwise $H > 1$ and it is expected to be quite small. Appendix 2 shows how to determine a standard error for $\hat{\Psi}_{mc}$ and $\hat{\Psi}_{sir}$ estimators. 

\textbf{\subsection{The choice of discrepancy functions}}

It is very common to use the \textit{sum of squared prediction errors} as a measure of discrepancy, because it is a form of cross-validation that provides a measure of model fit for those observations left out of the estimation procedure. \cite{alqgust2001} and \cite{Thall1997} adopt this measure for fitting a univariate dataset, described as $ r(\mathbf{s}_{y}, \mathbf{y}, \theta, \mathbf{y}^{rep}) = (\mathbf{y}^{rep}_{V[\mathbf{s}_{y}]} - \mathbf{y}_{V[\mathbf{s}_{y}]} )^2$.\\

We consider the \textit{mean squared prediction errors}. This measure can be written as

\begin{equation}\label{eq:mse}
r(\mathbf{s}_{y}, \mathbf{y}, \theta, \mathbf{y}^{rep}) = \frac{1}{n_V}\mid\mid (\mathbf{y}^{rep}_{V[\mathbf{s}_{y}]} - \mathbf{y}_{V[\mathbf{s}_{y}]}) \mid\mid ^2.
\end{equation}

This measure does not take into account the correlation between observations. As an alternative, we can consider the \textit{Mahalanobis distance}, which takes into account the covariance matrix of the common distribution of the two random vectors.

\small
\begin{equation}
r(\mathbf{s}_{y}, \mathbf{y}, \theta, \mathbf{y}^{rep}) = \sqrt {(\mathbf{y}^{rep}_{V[\mathbf{s}_{y}]} - \mathbf{y}_{V[\mathbf{s}_{y}]})'{\Sigma^{-1}}(\mathbf{y}^{rep}_{V[\mathbf{s}_{y}]} - \mathbf{y}_{V[\mathbf{s}_{y}]})} ,
\end{equation}

\normalsize

\noindent where $\Sigma = \tau^2 I_{\tau} + \sigma^2 R$ is covariance matrix of the regions formed by the locations that belong to both vectors. Therefore, using the Mahalanobis distance, we can compare de validation data sample, taking into account the spatial dependence.\\

Notice that so far we have considered a uniform prior for the validation sets which corresponds to assuming a homogeneous point process for the locations. In the following section, we propose a uniform prior in sub-regions to take into account spatial heterogeneity and to accommodate possible preferential sampling in the locations. \\

\section{Accounting for heterogeneity of locations\label{sec4.5}}

We consider stratified sampling to obtain training and validation sets. In stratified sampling, the region of $n$ units is first divided into sub-regions which are called \textit{strata} of $n_1, n_2, \ldots, n_{K}$ units, respectively. These sub-regions are non-overlapping, and together they comprise the whole region, so that, $ n= n_1+ n_2 + \ldots + n_K $. In each stratum, the full training data size $n_T$, is $n_{T1}+n_{T2}+\ldots+ n_{TK}$ and the full validation data size $n_V$  is $n_{V1}+n_{V2}+ \ldots n_{VK}$. If a simple random sample is taken in each stratum, the whole procedure is described as stratified random sampling. \\
 
Let $k$ denote the stratum and $T, V$ respectively denote the training and validation samples. The following symbols all refer to stratum $k$.

\begin{flushleft}
\begin{center}
\begin{tabular}{l }
\hline
\small \textbf{Notation}\\
\hline
\small $n$:   total number of sample spatial points in the study region \\
\small $n_k$: total number of spatial points in stratum $k$ \\
\small $n_{T_{k}}$:  number of spatial points of training data in stratum $k$ \\
\small $n_{V_{k}}$:  number of spatial points of validation data in stratum $k$ \\
\small$w_k=\frac{n_{V_{k}}}{n_V}$: stratum weight \\
\small $f_{V}= \frac{n_V}{n}$: sampling fraction, i.e., the ratio of \\
\small  validation sample size to  the total sample size.\\
\small $f_{T_{k}}= \frac{n_{T_{k}}}{n_k}$: training sampling fraction in the ${k}^{th}$ stratum  \\
\small $f_{V_{k}}= \frac{n_{V_{k}}}{n_k}$:  validation sampling fraction in the ${k}^{th}$ stratum  \\
\hline
\end{tabular}
\end{center}
\end{flushleft}

Stratification might produce a gain in precision in the estimates of characteristics of the whole region, if the variability inside each stratum is small and the variability between strata is large, \cite{Cochran1999}. It may be possible to divide a heterogeneous region into sub-regions, where each is internally homogeneous in the context of spatial cross-validation.\\

The following steps should be carried out to perform cross-validation using a stratified sampling scheme :

\begin{enumerate}

\item Stratify the study region into $k$ strata, where $n= n_1+ \ldots + n_K$, $k=1,2,\ldots,K$.   

\item In each stratum $k$, sampling via simple random sampling, to obtain the split vectors $\mathbf{s}^{(i,k)}_{y}$, where $k=1, \ldots, K$ represents the stratum and $i=1,2,\ldots, I_k$ the sizes of the split vectors generated in each stratum $k$. 
\end{enumerate}

The chosen sizes of the split vectors $I_k$ are the same in all strata, $I_k$, $k=1,2,\ldots,K$. Note that the sizes of split vectors $I_k$ do not need to be the same.\\

The generation of vector splits in each stratum $\mathbf{s}_{y}^{(1,k)}, \ldots, \mathbf{s}_{y}^{(I_{k},k)}$ is simulated jointly from $p(\mathbf{s}_{y})$. Thus, we define the vector $s_{\mathbf{y}}^{(i)}$ as the \textit{i-th} split vector considering all strata. 

$$
\mathbf{s}_{y}^{(i)} =  \left( \mathbf{s}_{y}^{(i,1)}, \mathbf{s}_{y}^{(i,2)}, \ldots, \mathbf{s}_{y}^{(i,K)} \right)_{1 \times n}, \hspace{0.2cm} i= 1, \ldots, I.  \\ \nonumber
$$

Notice that,
$$ \mathbf{s}_{y}^{(i,k)} = (s_1^{(i,k)}, s_2^{(i,k)}, \ldots, s_{n_k}^{(i,k)})_{1 \times n_k}.$$

The proposed stratification changes the sampling of spatial locations for validation and training sets, however, the sampling model is conditional on $\mathbf{s}_{y}$ and does not change with our proposal. Thus, the likelihood function is not affected. The vector of all observations can be written as

$$
\mathbf{y} = (y_{11}, \ldots, y_{1{n_1}}, \ldots, y_{ki}, \ldots, y_{K{n_K}})_{1 \times n}.
$$

\noindent The split vectors are jointly generated. In particular, the splits $\mathbf{s}_{y}$ are not uniformly distributed over the entire spatial domain. The proposed prior considers the stratification design and is given by

\begin{eqnarray}\label{priorst} \nonumber
p(\mathbf{s}_{y}) &=& {\dbinom{n_1}{n_{T_{1}}}}^{-1}{\dbinom{n_2}{n_{T_{2}}}}^{-1}\ldots \thinspace {\dbinom{n_K}{n_{T_{K}}}}^{-1} \\ 
&& \text{if} \quad \sum_{j=1}^{n_k} s_j^{(\cdot,k)} = n_{T_k},  \\  \nonumber
\end{eqnarray}

\noindent where each term of the product is the probability of choosing a sample of size $n_{T_{k}}$ in each stratum $k$. Our cross-validation assessments are based on the expectation under a \textit{reference distribution} of $(\mathbf{s}_{y}, \mathbf{y}, \theta, \mathbf{y}^{rep})$ as defined in \eqref{eq:eqreference}. We compute expectations with respect to the reference distribution for each stratum, denoted generically as 

\begin{equation}\label{eq:eq4.2}
{\Psi}_k = E\left\{ r_k(\mathbf{s}_{y}, \mathbf{y}, \theta, \mathbf{y}^{rep})\right\}, \quad k=1, \ldots, K,
\end{equation}

\noindent where the expression \eqref{eq:eq4.2} represents the expectation with respect to the reference distribution in each stratum $k$.

\textbf{\subsection{Stratified Estimators}}

To compute the stratified estimators, we jointly simulate the split vector $\mathbf{s}_{y}^{(1)}, \ldots, \mathbf{s}_{y}^{(I)}$ from $p(\mathbf{s}_{y})$ as defined in \eqref{priorst}. Following the same steps as in Section \ref{efficient}, the \textit{stratified MC estimator} can be obtained as  

\begin{eqnarray} \nonumber \label{eq:eq18}
{\hat{\Psi}^{st}_{mc}}  &=& \sum_{k=1}^{K} w_k \left\{ \frac{1}{I_k}\sum_{i=1}^{I_k}\frac{1}{J}\sum_{j=1}^{J} r_{k}(\mathbf{s}_{y}^{(i)}, \mathbf{y}, \theta_{ij}, y_{ij}^{rep})\right\} \\ \nonumber
    &=& \sum_{k=1}^{K}  w_k  \hat{\Psi}_{{mc}_{k}} \\ 
   &=&  \sum_{k=1}^{K} \hat{\Psi}^{st}_{{mc}_{k}}, \\ \nonumber
\end{eqnarray}

and the \textit{stratified SIR estimator} as,

\small
\begin{eqnarray} \nonumber \label{eq:eq19}
 {\hat{\Psi}_{sir}^{st}} &=& \sum_{k=1}^{K} w_k \left\{ \frac{1}{H} \sum_{h=1}^{H} \frac{1}{I_k} \sum_{i=1}^{I_k} \Psi_{hi}^{(k)} \right\} = \sum_{k=1}^{K} w_k  \hat{\Psi}_{{sir}_{k}}\\ 
    &=& \sum_{k=1}^{K}  \hat{\Psi}^{st}_{{sir}_{k}}, \\ \nonumber
\end{eqnarray}

\normalsize
\noindent where,

$$ \Psi_{hi}^{(k)} = \frac{\sum_{j=1}^{J} r_k(\mathbf{s}_{y}^{(i)}, \mathbf{y}, \theta_{hj}, y_{hj}^{rep}) w^{*}_{hj}}{\sum_{j=1}^{J} w^{*}_{hj}}, \hspace{0.2cm} k=1,\ldots, K, $$

\normalsize
\noindent and $w_k=\frac{n_{V_{k}}}{n_V}$ is the stratified weight, calculated using a proportional allocation of the sample $n_{V_k}$. Each weight term is given by $w^{*}_{hj}= p(\theta_{hj} \mid \mathbf{s}_{y}^{(i)}, \mathbf{y})/g(\theta_{hj})$. \\

\section{Simulated examples}\label{SecSimulate}

\begin{figure}[H]
\begin{center}
\begin{tabular}{ccc}
\includegraphics[width=0.3\textwidth]{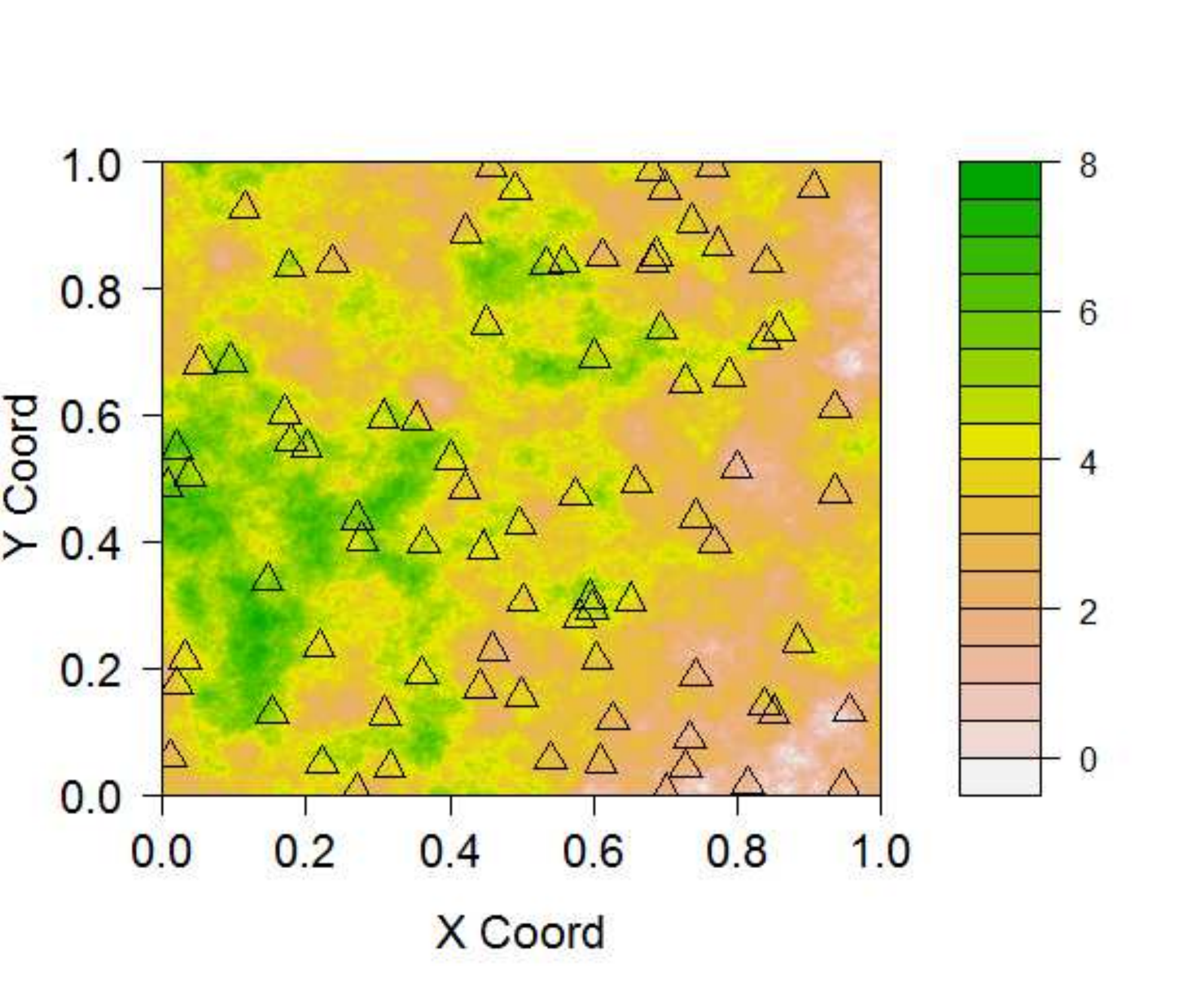}  & 
\includegraphics[width=0.3\textwidth]{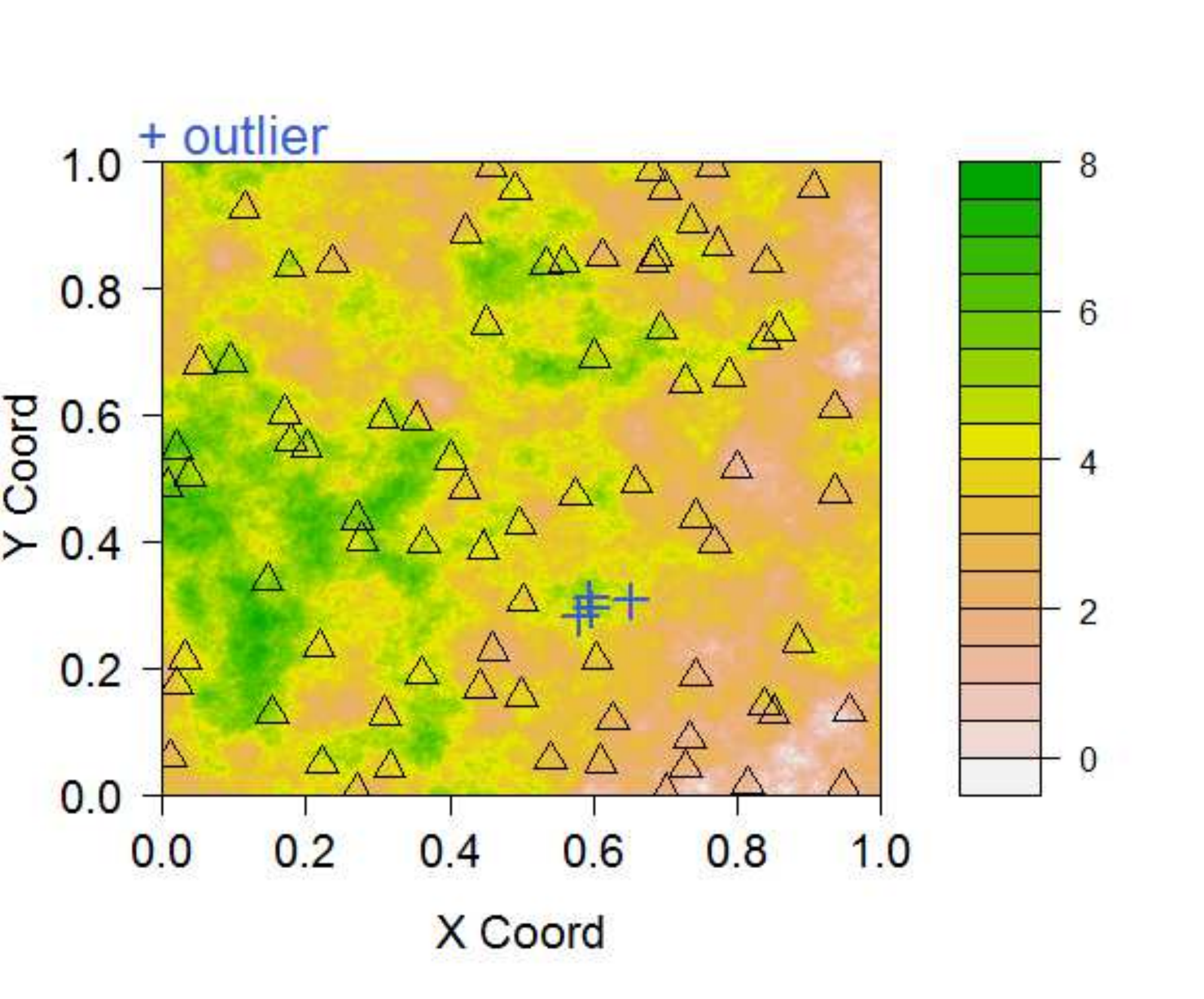} &   
\includegraphics[width=0.3\textwidth]{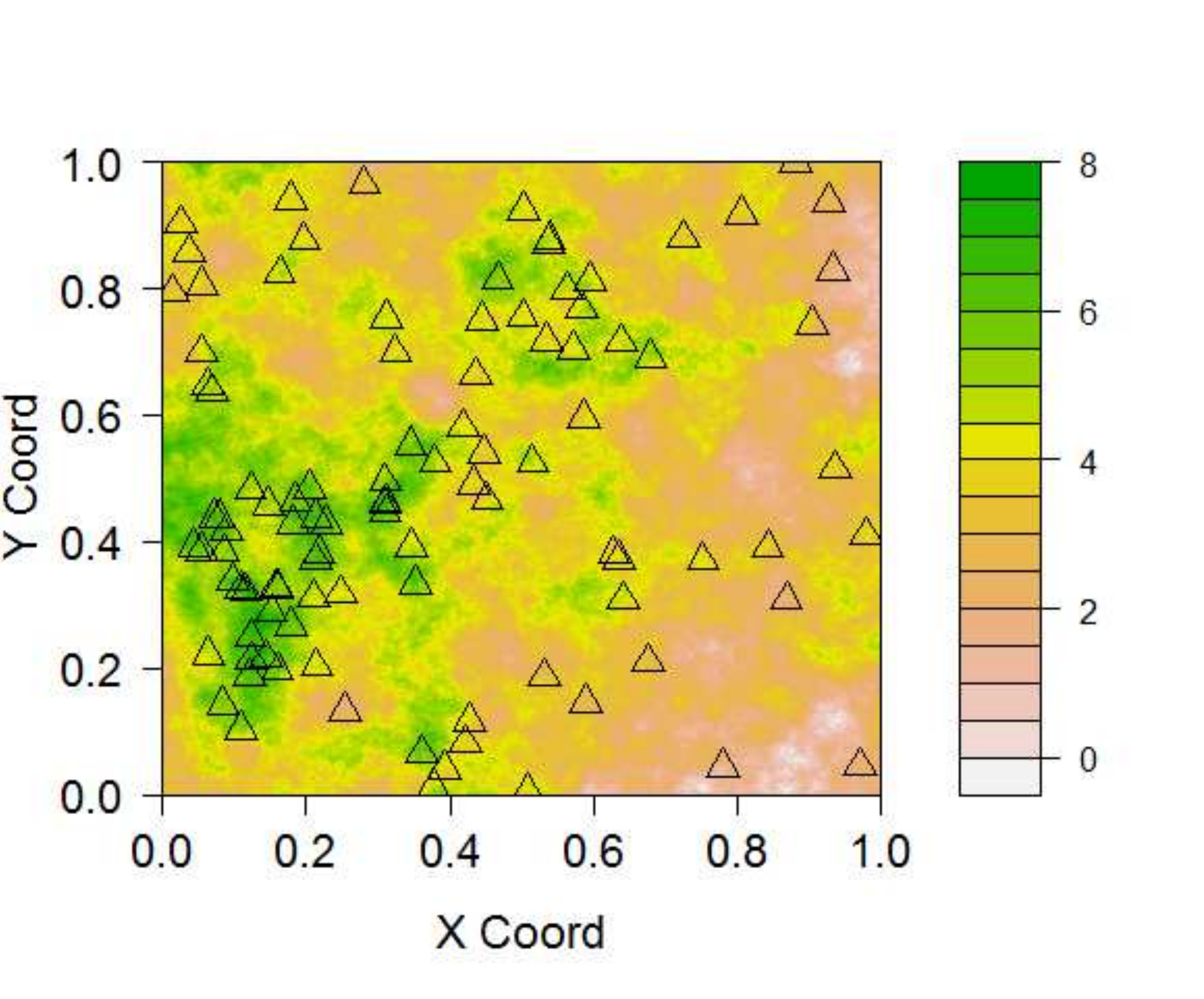} \\ 
\hspace{-0.8cm} (a) & \hspace{-0.6cm}(b) & \hspace{-0.6cm}(c) \\
\end{tabular}
\caption{\small Sample locations and underlying realizations of the signal process for the model used in the simulation study: (a) CRS ; (b) CRS with outlier; (c) preferential sample.}
\label{fig:fig3}
\end{center}
\end{figure}

To illustrate the usefulness of our cross-validation proposal we consider homogeneous and inhomogeneous processes under geostatistical modelling. We simulated data for different scenarios, considering different configurations for the location sampling. For each scenario, we first simulated an approximate realization of a stationary Gaussian process on the unit square, treating the spatially continuous process $S(\cdot)$ as constant within each lattice cell. After that, we simulated non-preferentially or preferentially scenarios according to each of the sampling designs presented in Figure \ref{fig:fig3}. The data were generated from equation \eqref{eq:sec2eq1} where:

\begin{description}
\item \textbf{(A1)} $S$ is stationary Gaussian process with mean 0, variance $\sigma^2$ and correlation function $\rho({u},\phi)=$ \\$Corr(S(x), S(x'))$ for any $x$ and $x'$ from a distance ${u}$ apart.
\item \textbf{(A2)} $X \mid S$ is an inhomogeneous Poisson process with log-linear intensity function
\begin{equation}\label{eq:sec2eq2}
\lambda(x) = exp\left\{\alpha + \beta S(x) \right\}
\end{equation}
\item \textbf{(A3)}  $Y \mid S,X$ is a set of mutually independent Gaussian variates with $Y_i \sim N(\mu + S(x_i), \tau^2)$.
\end{description}

Notice that if $\beta=0$, the sampling is done completely at random (CRS), i.e., a homogeneous Poisson process. The simulated surface in (A1) is given by a Gaussian process with the following parameters: $\mu=4, \sigma^2 = 1.5,\phi=0.15,\kappa=0.5$ and $\tau^2=0.25$. We adopted the exponential correlation function in all scenarios.\\

Scenario 1 -- CRS, we account for the case representing the situation where the $\beta$ parameter leads to null intensity function $\lambda(x)$. Therefore, the point pattern does not rely on $S$ and because $\lambda$ is a constant, we have a completely random point pattern. A dataset was simulated with sample size equal to $n=82$ and intensity parameters $\beta=0, \alpha=4.605$. This is presented in Figure \ref{fig:fig3} (a).\\

Scenario 2 -- CRS with outlier, we study the same surface of Figure \ref{fig:fig3} (a) with observations contaminated by summing a random increment $u \sigma$, such that $\sigma$ is the observational standard deviation and $u \sim U(6,8)$ for observations 10, 48, 50 and 82. The contaminated locations considered are neighbours in space. This is presented in Figure \ref{fig:fig3} (b).\\

Scenario 3 -- preferential sampling, we choose the configuration with highest concentration of points in a given region. The spatial locations are simulated from the non-homogeneous process. This process represents the inhomogeneous Poisson process, with intensity $\lambda(x)$, $\alpha= 2.996$, $\beta=1.0$ and $n=100$. This is presented in Figure \ref{fig:fig3} (c).\\

For all sample designs presented above, we made a cross-validation comparison of the models. In the following, it is considered a mixture model in a geostatistical context. The general spatial mixture model considered for model fitting is given by

\begin{equation}\label{NGprocess}
Y(x)  = \mu + \frac{S(x)}{\delta^{1/2}(x)} + Z(x), \hspace{0.3cm} \forall \thinspace x \in A.
\end{equation}

Consider the observations $y_1, \ldots, y_n$ at locations  $x_1, \ldots, x_n$.The particular cases of model \eqref{NGprocess} investigated in this work are detailed as follows. \\

\noindent  {\bf (M1) Gaussian model:} We set $\delta(x)=1$, $\forall x \in A$ as a benchmark. The distribution of $\mathbf{Y}$ is
\begin{equation}\label{eq:sec2eq2a}
\mathbf{y} \mid \mu, \sigma^2, \phi \sim N \left( \mathbf{1} \mu, \tau^2 I_n + \sigma^2 R \right).
\end{equation}

\noindent  {\bf (M2) Student-t model:} Define $\delta(x)=\delta$, $\forall x \in A$ such that $\delta \mid \nu \sim Ga(\nu/2,\nu/2)$. Then, by marginalization, the distribution of $\mathbf{Y}$ is
\begin{equation}
\label{eq:eq12}
\mathbf{y} \mid \mu, \sigma^2,\phi,\nu \sim ST \left(\mathbf{1}\mu, \nu, \tau^2 I_n + \sigma^2 R \right).  
\end{equation}

Similar to the Gaussian process, the Student-t process has the advantage of depending on the mean and covariance functions. Details about the Student-t process in a non-Bayesian context may be seen in \cite{Omre06}. \\

\noindent  {\bf (M3) Gaussian-Log-Gaussian model:}  As proposed by \cite{PSteel06}, this process is able to capture heterogeneity in space through a mixing process used to increase the Gaussian process variability, 

\begin{equation} \label{eq:eq12a}
\mathbf{y} \mid \mu,\sigma^2,\phi, \mathbf{\Delta}  \sim N \left(\mathbf{1} \mu,  \tau^2 I_n + \sigma^2 ({\mathbf{\Delta}^{-1/2}} \thinspace R \thinspace {\mathbf{\Delta}^{-1/2}}) \right).
\end{equation}

\noindent This model assumes $\boldsymbol{\Delta}= diag(\delta(x_1), \ldots, \delta(x_n))$ and $ln(\boldsymbol{\delta}) \sim Normal_n \left(-\frac{\upsilon}{2}, \upsilon \thinspace R \right)$. This mixing generates a multivariate scale mixture of Normals. 
Properties, estimation and prediction for the GLG model are introduced in \cite{PSteel06} and extended to the space-time case in \cite{Fons11}. The $\upsilon \in \mathbb{R}^{+}$ is a scalar parameter introduced into the distribution $ln(\boldsymbol{\delta})$ and variation inflation is achieved when it is close to zero.\\

For the CRS and CRS with outlier scenarios, we arbitrarily choose $n_T= 77, n_V=5$. The sampling process in space was done randomly without considering the correlation between spatial locations, that is, $p(\mathbf{s}_{y})$ is assumed to be a uniform distribution. The parameter $\nu$ is fixed at 3 for M2 in the CRS scenario. MC and SIR estimators are based on averaging over the same $I=100$ splits. We sampled from the posterior of the model parameters using Metropolis-Hastings with random walk proposals, which led to reasonable acceptance rates in the vicinity of 30\% to 50\% for each parameter. In this case, we fixed the nugget effect $\tau^2$ at 0.25.\\

For the preferential scenario, we considered $n_T= 95$ and $n_V=5$. The training and validation sampling process in space were done randomly without considering the correlation between spatial locations. We obtained posterior samples of this model parameters via the Metropolis-Hastings with random walk proposals, which led to reasonable acceptance rates in the vicinity of 30\% to 55\% for each parameter. In this case, it is adopted the nugget effect $\tau^2$ equal to 0.25. \\

As can be seen in Table \ref{tabcomp1}, the execution time for the SIR estimator using $H=5$ is shorter than that for the  MC estimator. The high computational cost is due to the need to calculate the covariance matrix for each split vector sampled using a machine Intel Core i5 3210M CPU 2.50GHz 2.50GHz 4GB RAM memory - System 64 bits - Windows 10 Home. \\

Table \ref{tab1} presents the mean square prediction error measures based on the MC and SIR estimators with their respective standard errors. It is clear that the SIR estimator variability must be greater than that of the MC estimator, because the estimator is a heuristic approximation based on the same amount of data. However, the SIR estimator is a good approximation of the original estimator. \\

Table \ref{tab1}(CRS) presents original estimates of mean square error. M3 has similar performance as that of M1. This is due to the fact that M1 is a particular case of M3. In particular, when $\upsilon \rightarrow 0 $, we retrieve Normal tails, while larger values of $\upsilon$ induce thicker tails. In the Gaussian case, $\boldsymbol{\delta}$ is equal to $\mathbf{1}$. M2 has much worse performance. \\

\begin{table}[H]
\begin{center}
\caption{Computational Time (hours)}
\label{tabcomp1}
\scalebox{0.9}{
\begin{tabular}{lcccccc}
\hline\noalign{\smallskip}
&  \multicolumn{2}{ c }{M1} & \multicolumn{2}{ c }{M2} & \multicolumn{2}{ c }{M3} \\ 
&  MC & SIR & MC & SIR  & MC & SIR  \\ 
\noalign{\smallskip}\hline\noalign{\smallskip}
CRS &   11.20 &  2.32 & 15.44   & 2.33   &  33.80 & 3.50 \\
CRS with outlier & 11.20 & 2.00  & 13.80   &  3.12  & 20.61 & 3.48 \\
Preferential &16.12 & 2.73  & 23.72 &  3.05  &  41.36 & 4.98   \\ 
\noalign{\smallskip}\hline
\end{tabular}}
\end{center}
\end{table}

\begin{table}[H]
\begin{center}
\caption{Cross-validation using mean square error for M1, M2 and M3 models in each scenario. The same splits are considered for all models.}
\label{tab1}
\begin{tabular}{ l ll l l }
\hline\noalign{\smallskip}
& &  \multicolumn{1}{ c }{M1} & \multicolumn{1}{ c }{M2} & \multicolumn{1}{ c }{M3} \\ 
\noalign{\smallskip}\hline\noalign{\smallskip}
CRS & MC & 3.496 ($1.1\times 10^{-5}$)  & 7.210 (0.009)   &  3.090 ($9.6\times 10^{-6}$)\\
& SIR &3.122 (0.003) &  6.686 (0.085) & 2.833 (0.001)   \\ 
\noalign{\smallskip}
CRS with outlier & MC & 10.559 ($1.2\times 10^{-4}$) & 22.572 (0.022)  & 8.371 ($1.1\times 10^{-4}$) \\
& SIR & 10.487 (0.150) & 23.247 (0.773)  &  7.049 (0.121)  \\ 
\noalign{\smallskip}
Preferential & MC & 5.175 ($2.6\times 10^{-5}$)&  7.624 (0.002)  &  5.075 ($5.5\times 10^{-5}$) \\
& SIR & 5.017 (0.009)&  9.009 (0.311) &  4.683 (0.015)   \\
\noalign{\smallskip}\hline
\end{tabular}
\end{center}
\end{table}

Although the Student-t process has heavier tails than the Gaussian, it does not have the flexibility to model georeferenced data. Its behaviour is inadequate when compared to M1 and M3. According to \cite{Lobo2016}, the Student-t process inflates the variance of the whole process in the presence of outliers and does not allow for both individual or regional outlier detection and different kurtosis behaviours across space.\\

Table \ref{tab1} presents the results of the three proposed models. Observe that estimates are more precise for the GLG model. This is due to the fact that this model tends to detect sub-regions with larger variability. The use of M3 scale mixing reduces the variance estimate. M1 and M2 overestimate the variance.\\

In the case of Gaussian data contaminated by outliers, M3 has smaller discrepances. The Gaussian Log Gaussian process is able to capture heterogenity in space through a mixing process used to increase the Gaussian process variability. This proposal is an alternative to the usual Gaussian model which is very sensitive to outliers. 

\textbf{\subsection{Analysing heterogeneity of locations}}

In stratified sampling we have to sample in all strata of the spatial region \textit{A}. This condition does not occur in random sampling. Stratification is carried out in order to not favour any particular area, making the process as uniform as possible. After that, the sampling procedure is done randomly in each unit. \\

\begin{figure}[H]
\begin{center}
\begin{tabular}{ccc}
\includegraphics[width=0.3\textwidth]{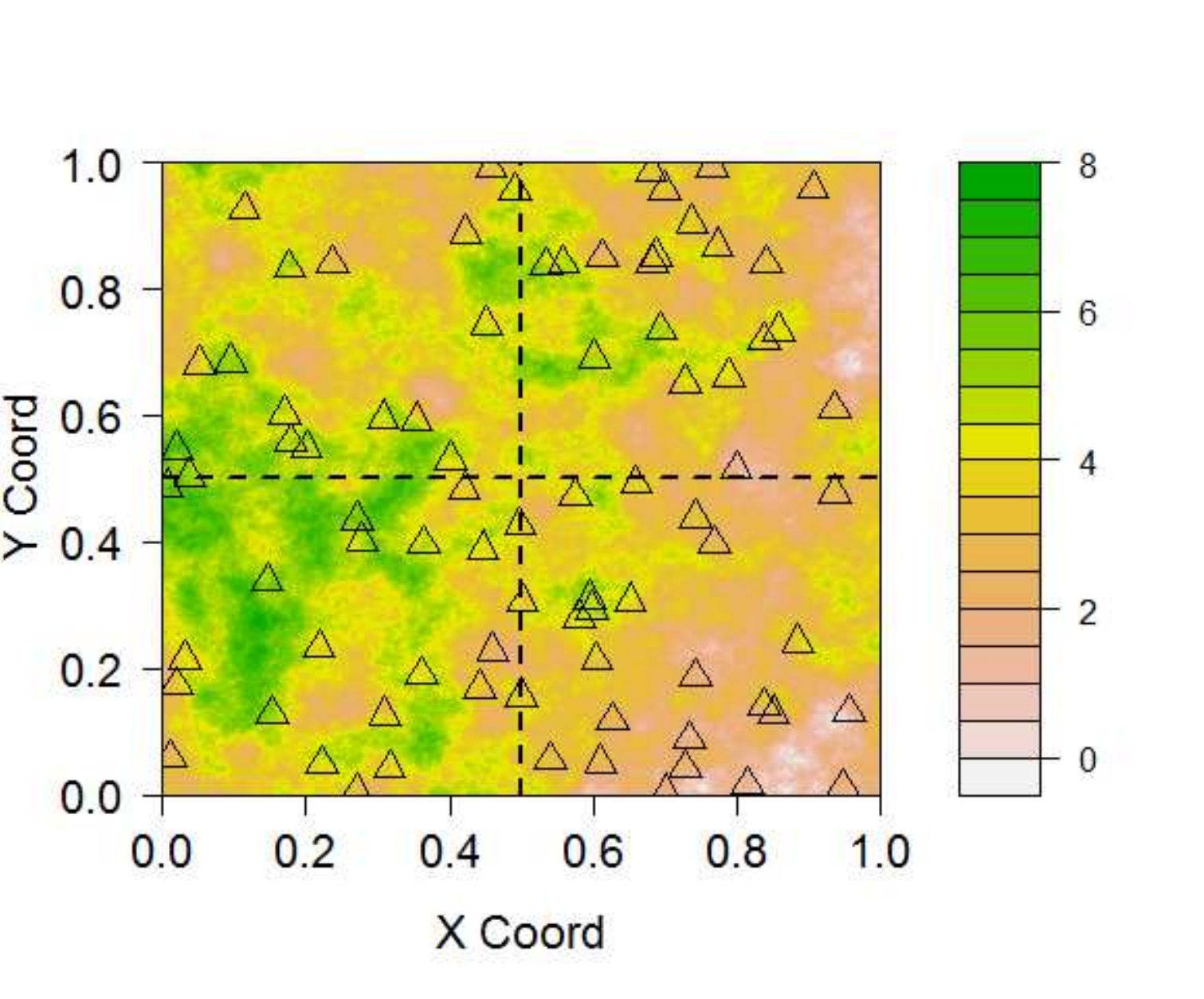}  & 
\includegraphics[width=0.3\textwidth]{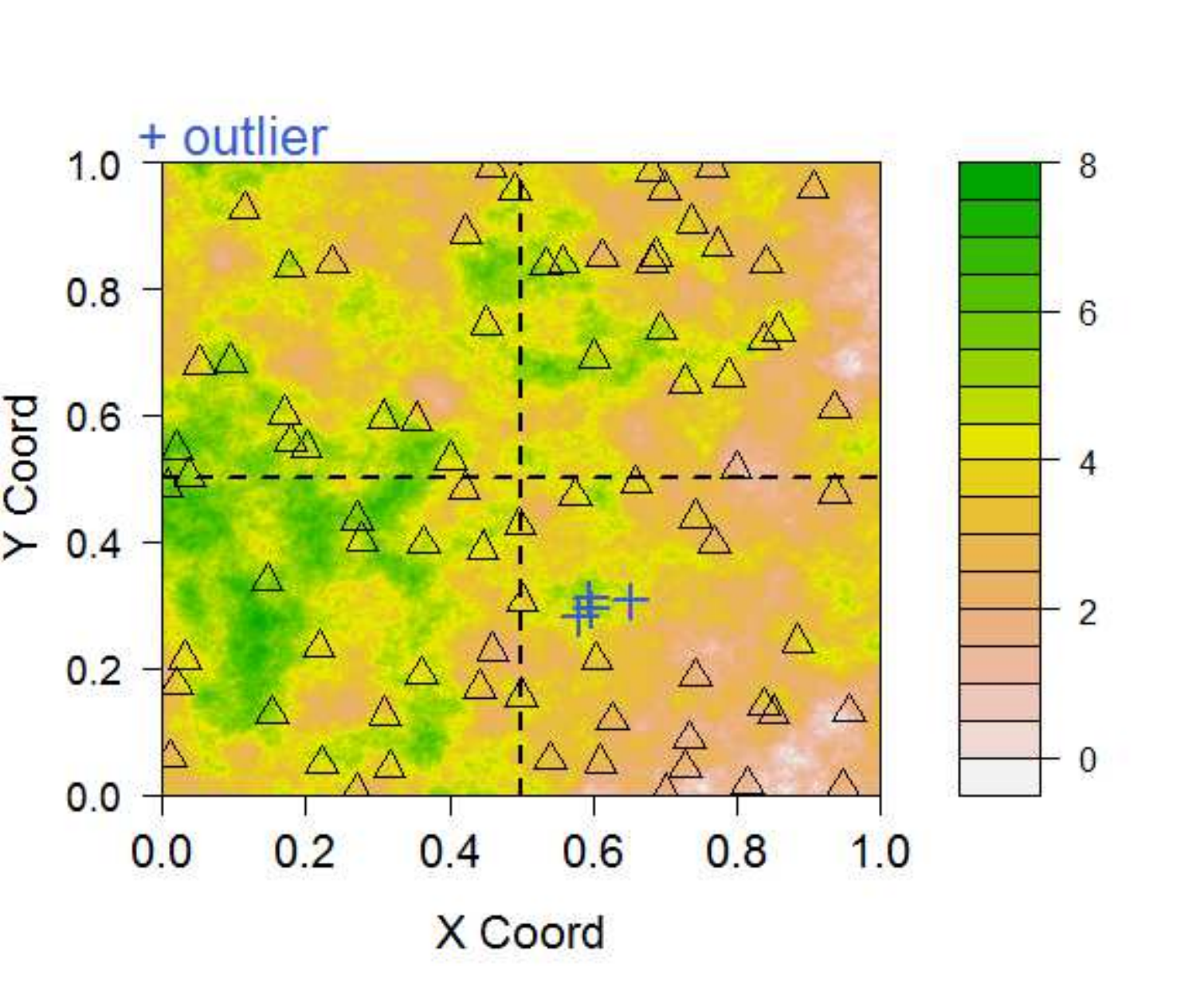} &   
\includegraphics[width=0.3\textwidth]{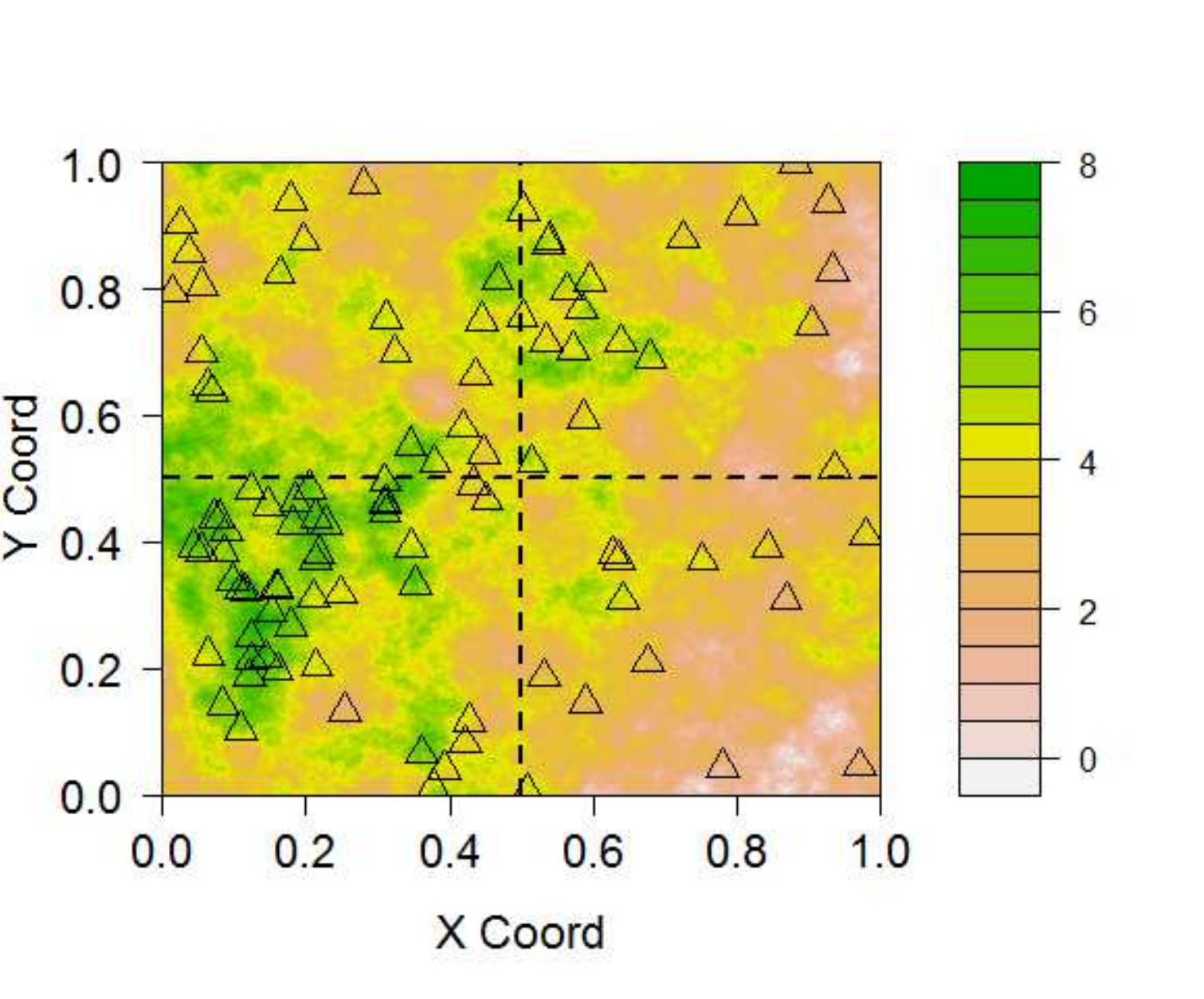} \\ 
\hspace{-0.8cm} (a) & \hspace{-0.6cm}(b) & \hspace{-0.6cm}(c) \\
\end{tabular}
\caption{\small Sample locations and underlying realizations of the signal process for the model used in the simulation study: (a) CRS ; (b) CRS with outlier; (c) preferential sample.}
\label{fig:fig4}
\end{center}
\end{figure}

We stratified the unit square into four strata for all scenarios. Figure \ref{fig:fig4} presents the stratification of the study region in $A$. Since we have a homogeneous process, we expected the number of events to be similar in each stratum, as shown in Table \ref{tab4} (CRS). The choice of $I_k$ is arbitrary for each stratum.\\

\begin{table}[H]
\begin{center}
\caption{Stratified design for all scenarios.}
\label{tab4}
\begin{tabular}{c}
\scalebox{1.0}{\begin{tabular}{c cccc}
  \hline\noalign{\smallskip}
  & \multicolumn{4}{c}{CRS / outlier} \\
  \textit{strata} & $n_k$ & $n_{Tk}$ & $n_{Vk}$ & $w_k$   \\ 
  \noalign{\smallskip}\hline\noalign{\smallskip}
  {1}& 21 & 19 & 2 &   0.250  \\
  {2}& 17 & 15 & 2  & 0.250 \\
  {3}& 24 & 22 & 2 &  0.250\\
  {4}& 20 & 18 & 2  & 0.250 \\
  \noalign{\smallskip}\hline
  \textit{total} & 82 & 74 & 8 & 1 \\
  \noalign{\smallskip}\hline
  \end{tabular}}

\\
\\

\scalebox{1.0}{\begin{tabular}{c cccc}
  \hline\noalign{\smallskip}
  & \multicolumn{4}{c}{Preferential} \\
  \textit{strata} & $n_k$ & $n_{Tk}$ & $n_{Vk}$ & $w_k$   \\ 
  \noalign{\smallskip}\hline\noalign{\smallskip}
  {1}& 47 &42  & 5 &   0.500 \\
  {2}&20  &18& 2&   0.200\\
  {3}& 13  &12&1  & 0.100\\
  {4}& 20&18& 2 &  0.200\\
  \noalign{\smallskip}\hline
  \textit{total} & 100 & 90 & 10 & 1 \\
  \noalign{\smallskip}\hline
  \end{tabular}}
\end{tabular}
\end{center}
\end{table}

Table \ref{tab4} shows the stratification and selection of training and validation data for the respective strata via the sampling process for all scenarios. Again, the execution time for the SIR estimator using $H=5$ is shorter than of the MC estimator, which can be seen in Table \ref{tabcomp2}.

\begin{table}[H]
\begin{center}
\caption{Computational Time (hours)}
\label{tabcomp2}
\scalebox{0.8}{
  \begin{tabular}{l cccccc}
  \hline\noalign{\smallskip}
  &  \multicolumn{2}{ c }{M1} & \multicolumn{2}{ c }{M2} & \multicolumn{2}{ c}{M3} \\ 
  &  MC & SIR & MC & SIR & MC & SIR \\ 
  \noalign{\smallskip}\hline\noalign{\smallskip}
  CRS & 14.67 &3.42  & 13.48 & 3.28 & 37.66 & 4.71 \\
  CRS with outlier & 13.48 & 3.05 &  14.36  &  2.86   & 25.55 &   4.78\\
  Preferential & 19.92 &3.46  & 18.36   & 4.14   &  30.387 & 3.48   \\
  \noalign{\smallskip}\hline
  \end{tabular}}
\end{center}
\end{table}

The hypothesis of \textit{complete spatial randomness} establishes that the number of events per unit area is a constant, $\lambda$,  over all the region considered. In the homogeneous case, we have approximately the same estimates in each stratum. The stratified estimator is equivalent to the global estimator presented in Table \ref{tab5} (CRS), considering the mean square error as the measure of precision. This occurs because the sample size is similar in all strata. Regarding the performance of the stratified estimator, their variability is smaller, mainly for the SIR estimator. Stratification reduces the variability of the process in order to transform the stratum as homogeneously as possible. This can be confirmed by analysing Tables \ref{tab5} and \ref{tab5b} for all scenarios. \\

Clearly there is an increase in the accuracy of the estimator by stratifying the spatial region. Furthermore, the stratification allows the identification of lack of fit for all models in region 3 for the scenario with outliers. All models have much larger values of the discrepancy function for stratum 3, which contains the contaminated observations, as presented in Tables \ref{tab5} and \ref{tab5b} (CRS with outlier) using different discrepancy measures .\\

Note, however the reduction of the variability of both estimators. The performance of the models is similar to that of Section 6 except for M3 in the preferential scenario. In this case, the GLG model has much better performance, indicating that dividing the region in sub-regions provides a much better predictive performance of this model in all sub-regions.\\

M3 again stands out because of its ability to capture heterogeneity in space.  This is an appealing feature in the non-homogeneous setup, because strata with a high concentration of events tend to present larger variability.\\

The use of Mahalanobis distance discrepancy improved model discrimination by indicating the Gaussian model as the best model in scenario 1. This is an expected result, since the data are generated by the Gaussian model. On the other hand, when we consider the mean squared error as a measure, the Gaussian and GLG models have approximately the same results, with a bit better performance of the GLG model. This can be seen in Tables \ref{tab5} and \ref{tab5b}.

\begin{landscape}

\begin{table}[H]
\begin{center}
\caption{Stratified cross-validation using mean square error for M1, M2 and M3 models in each scenario. The same splits are considered for all models. }
\label{tab5}
\scalebox{0.8}{
  \begin{tabular}{ l l cc cc cc}
  \hline\noalign{\smallskip}
  & & \multicolumn{2}{c}{M1} & \multicolumn{2}{c}{M2}  &  \multicolumn{2}{c}{M3}\\
  & \textit{strata} & MC  &  SIR  & MC  & SIR  & MC  & SIR  \\
  \noalign{\smallskip}\hline\noalign{\smallskip}
  CRS &1 & 3.617 ($ 2.2\times 10^{-5}$)   & 3.804 ($ 1.1\times 10^{-3}$)  &  7.081 (0.002) &  6.533 (0.003)  & 3.033 ($ 1.5\times 10^{-5}$)     & 2.311 (0.002)  \\  
  &2 & 3.595 ($ 2.2 \times 10^{-5}$) & 3.881 ($ 2.7\times 10^{-3}$) &  7.319 (0.028)  &5.770  (0.004)   & 3.160 ($ 1.6 \times 10^{-5}$)   & 2.415 (0.003)   \\ 
  &3 & 3.439 ($ 2.1 \times 10^{-5}$)&   3.546 ($8.0 \times 10^{-4}$) &  6.919 (0.003) &  6.095 (0.003)  & 2.743 ($ 1.4 \times 10^{-5}$)   &  2.089 (0.003) \\
  &4 & 3.255 ($ 2.1 \times 10^{-5}$)&   3.486 ($1.1 \times 10^{-3}$) & 6.925 (0.005)  & 5.726 (0.008) & 2.665 ($ 1.4 \times 10^{-5}$) & 1.900 (0.002) \\
  \noalign{\smallskip}\hline
  & $\hat{\Psi}^{st}$ & $\mathbf{3.477}$ ($ 5.4 \times 10^{-6}$) & $\mathbf{3.679}$  ($3.6 \times 10^{-4}$)  &  $\mathbf{7.061}$ (0.002) & $\mathbf{6.031}$ (0.001)  &  $\mathbf{2.900}$ ($5.9 \times 10^{-5}$)  &  $\mathbf{2.179}$ ($6.4 \times 10^{-4}$)\\
  \noalign{\smallskip}\hline
  outlier &1 &   6.173 ($6.9 \times 10^{-5}$) &  6.747 (0.023)    &     17.756 (0.019)    & 22.570 (0.013)   &   2.808 ($1.5 \times 10^{-5}$)
  & 2.152 (0.002) \\    
  &2 & 6.843 ($8.2 \times 10^{-5}$)   &  6.804 (0.025)     &   19.549 (0.289)    & 21.608 (0.013)   &  3.367 $(1.9 \times 10^{-5}$)  &   2.776 (0.002) \\ 
  &3 & 22.071  ($1.0 \times 10^{-3}$) &  22.738 (0.215)   &   33.636 (0.030)    &  36.981 (0.218)     &  18.996 ($8.8 \times 10^{-4}$)  & 17.728 (0.272) \\
  &4 &  5.873 ($6.6 \times 10^{-5}$)  &  6.599 (0.030)   &   17.871 (0.043)     & 21.920 (0.017) &   2.516 ($1.4 \times 10^{-5}$) &  1.947 (0.002)\\
  \noalign{\smallskip}\hline
  &$\hat{\Psi}^{st}$  & $\mathbf{10.240}$ ($8.0 \times 10^{-5}$)   & $\mathbf{10.722}$ (0.0183)  & $\mathbf{22.203}$ (0.024) &  $\mathbf{25.769}$ (0.016)& $\mathbf{6.922}$ ($9.3 \times 10^{-4}$)   &  $\mathbf{6.151}$ (0.017) \\
  \noalign{\smallskip}\hline
  preferential &1 & 7.338($ 5.4 \times 10^{-5}$)&  7.402($3.7 \times 10^{-4}$) &  9.986(0.005)  & 11.398(0.013)  & 4.456 ($ 1.1\times 10^{-5}$)  & 5.032  (0.002)\\
  &2 & 3.342 ($ 2.4 \times 10^{-5}$)& 3.366 ($2.4 \times 10^{-5}$) &   5.691 (0.003)  &  6.206 (0.083)  &  3.307 ($ 1.8 \times 10^{-5}$)   &  2.352 (0.002) \\ 
  &3 & 3.198 ($ 3.5 \times 10^{-5}$)& 3.314  ($9.2 \times 10^{-5}$) &  5.453 (0.002)  & 8.739 (0.013) & 3.751 ($ 4.1 \times 10^{-5}$)   & 2.377  (0.003)\\
  &4 & 3.670 ($ 2.8 \times 10^{-5}$)&  3.833 ($2.9 \times 10^{-5}$)&   6.243 (0.015)  &  6.989 (0.083)  &  3.448 ($ 1.9 \times 10^{-5}$) & 2.510  (0.002)\\
  \noalign{\smallskip}\hline
  &$\hat{\Psi}^{st}$  &  $\mathbf{5.391}$ ($ 1.6 \times 10^{-5}$)& $ \mathbf{5.472}$ ($ 4.3 \times 10^{-4}$)  &  $\mathbf{7.925}$ (0.002) & $\mathbf{8.333}$ (0.011) & $\mathbf{3.954}$ ($ 8.8 \times 10^{-5}$) & $\mathbf{3.726}$ ($6.0 \times 10^{-4}$) \\
  \noalign{\smallskip}\hline
  \end{tabular}}
\end{center}
\end{table}

\begin{table}[H]
\begin{center}
\caption{Stratified cross-validation using Mahalanobis distance for M1, M2 and M3 models in each scenario. The same splits are considered for all models. }
\label{tab5b}
\scalebox{0.8}{
  \begin{tabular}{ l l cc cc cc}
  \hline\noalign{\smallskip}
  & & \multicolumn{2}{c}{M1} & \multicolumn{2}{c}{M2}  &  \multicolumn{2}{c}{M3}\\
  & \textit{strata} & MC  &  SIR  & MC  & SIR  & MC  & SIR  \\
  \noalign{\smallskip}\hline\noalign{\smallskip}
  CRS &1 &   1.806 ($ 8.0 \times 10^{-4}$)  &  1.812 ($ 1.1 \times 10^{-4}$)  &  3.426 ($4.6 \times 10^{-6}$) & 3.312 ($1.1 \times 10^{-3}$)   &   2.056 ($ 1.8 \times 10^{-6}$) & 2.080 ($1.5 \times 10^{-4}$)  \\  
  &2 &  1.781 ($ 1.2 \times 10^{-6}$)   & 1.789 ($ 1.5 \times 10^{-4}$)    & 3.389 ($5.0 \times 10^{-6}$)  &  3.256 ($1.5 \times 10^{-4}$)    &   2.091 ($1.9 \times 10^{-6}$)   &  2.038 ($ 4.2 \times 10^{-4}$)\\  
  &3 &   1.725 ($ 1.1 \times 10^{-6}$) & 1.693 ($ 5.8 \times 10^{-5}$)    &  3.340 ($4.6 \times 10^{-6}$)   &   3.324 ($1.3 \times 10^{-4}$)   & 1.919 ($1.6 \times 10^{-6}$)   &   1.961 ($ 4.0 \times 10^{-4}$)\\  
  &4 &   1.663 ($ 1.2 \times 10^{-6}$)  & 1.672 ($ 9.9 \times 10^{-5}$) & 3.269 ($4.8 \times 10^{-6}$)  &    3.164 ($2.5 \times 10^{-4}$)  &  1.854 ($1.7 \times 10^{-6}$)    &  1.905 ($ 5.4 \times 10^{-4}$)\\  
  \noalign{\smallskip}\hline
  & $\hat{\Psi}^{st}$ &  $\mathbf{1.744}$ & $\mathbf{1.741}$ ($ 2.7 \times 10^{-5}$) &  $\mathbf{3.356}$ ($ 1.2 \times 10^{-6}$) & $\mathbf{3.264}$ ($9.9 \times 10^{-5}$)   & $\mathbf{1.980}$ ($ 4.5 \times 10^{-7}$)  &  $\mathbf{1.996}$ ($ 1.3 \times 10^{-4}$) \\
  \noalign{\smallskip}\hline
  outlier &1 & 2.406 ($3.9 \times 10^{-6}$)    & 2.138 ($4.2 \times 10^{-4}$)  & 3.914 ($9.8 \times 10^{-6}$)  &  3.672 ($1.2 \times 10^{-3}$) & 1.979 ($3.9\times 10^{-6}$)   & 1.519 ($1.9 \times 10^{-4}$)  \\  
  &2 &  2.428 ($ 4.1 \times 10^{-6}$)   &  2.480 ($5.5 \times 10^{-4}$)& 4.024 ($1.0 \times 10^{-5}$)  & 4.156 ($1.8 \times 10^{-3}$)   &  1.199 ($4.1 \times 10^{-6}$)   &  1.555 ($2.3 \times 10^{-4}$)\\  
  &3 &  3.401 ($ 6.8 \times 10^{-6}$)   &   5.295  ($1.0 \times 10^{-2}$)  &  6.154 ($2.2 \times 10^{-5}$) & 5.906 ($4.3 \times 10^{-3}$)    &  2.671 ($6.8 \times 10^{-6}$)  &  2.683 ($9.8 \times 10^{-4}$) \\  
  &4 &   2.324 ($ 3.7 \times 10^{-6}$)  & 1.969  ($4.2 \times 10^{-4}$) & 3.830 ($9.8 \times 10^{-6}$)  &   3.742 ($1.3 \times 10^{-3}$)  &   1.825 ($3.7 \times 10^{-6}$) &  1.471 ($2.2 \times 10^{-4}$)\\  
  \noalign{\smallskip}\hline
  &$\hat{\Psi}^{st}$  &  $\mathbf{2.640}$ ($ 1.7 \times 10^{-6}$)  & $\mathbf{2.970}$  ($7.5 \times 10^{-4}$)  & $\mathbf{4.483}$ ($3.2 \times 10^{-6}$)  &  $\mathbf{4.368}$ ($5.4 \times 10^{-4}$)  &  $\mathbf{1.918}$ ($1.8 \times 10^{-5}$)  & $\mathbf{1.807}$ ($1.0 \times 10^{-4}$) \\  
  \noalign{\smallskip}\hline
  preferential &1& 3.571 ($1.7 \times 10^{-6}$)   &  3.632 ($1.0 \times 10^{-4}$) & 4.924 ($9.1 \times 10^{-6}$)  &  4.988 ($5.9 \times 10^{-4}$)   &   3.380 ($1.5 \times 10^{-6}$) & 3.165 ($6.6 \times 10^{-4}$)   \\  
  &2 &  1.560 ($9.9 \times 10^{-6}$)   & 1.581 ($3.9 \times 10^{-4}$) &  2.217 ($4.6 \times 10^{-6}$) &  2.078 ($3.4 \times 10^{-4}$)  & 1.680 ($1.2 \times 10^{-6}$)    &   1.838 ($4.2 \times 10^{-4}$)   \\  
  &3&  0.969 ($7.7 \times 10^{-7}$)  &  0.950 ($1.7 \times 10^{-4}$) & 1.369 ($2.9 \times 10^{-6}$)  &  1.336 ($5.9 \times 10^{-4}$)   & 1.164 ($1.0 \times 10^{-6}$)    &  1.153 ($3.3 \times 10^{-4}$)   \\  
  &4 &   1.641 ($1.0 \times 10^{-6}$) &  1.646 ($3.7 \times 10^{-4}$)   &  2.327 ($4.6 \times 10^{-6}$) &  2.251 ($6.3 \times 10^{-4}$)   &   1.726 ($1.2 \times 10^{-6}$) &  1.953 ($4.4 \times 10^{-4}$)  \\  
  \noalign{\smallskip}\hline
  &$\hat{\Psi}^{st}$  &  $\mathbf{2.523}$ ($2.8 \times 10^{-7}$)   & $\mathbf{2.557}$ ($1.2 \times 10^{-4}$)   & $\mathbf{3.508}$ ($1.3 \times 10^{-6}$)   &   $\mathbf{3.494}$ ($1.3 \times 10^{-4}$) & $\mathbf{2.488}$ ($5.0 \times 10^{-6}$)   & $\mathbf{2.456}$ ($1.1 \times 10^{-4}$) \\  
  \noalign{\smallskip}\hline
  \end{tabular}}
\end{center}
\end{table}

\end{landscape}


\section{Application to a rainfall data} 

The dataset used in this application contains the total rainfall (in \textit{mm}) recorded in  Octorber 2010 in 31 locations in the city of Rio de Janeiro, Brazil, obtained from \textit{Instituto Pereira Passos}, known for offering one of the largest collections of maps and statistical data of Rio de Janeiro available in the \textit{Armazem de Dados}. Stations with missing information were removed from the study. \cite{Gustavo2015} analyzed the same kind of data for October 2005 in the context of achieving the optimal design using preferential sampling. \\

\begin{figure}[H]
\begin{center}
\begin{tabular}{c}
\includegraphics[width=0.85\textwidth]{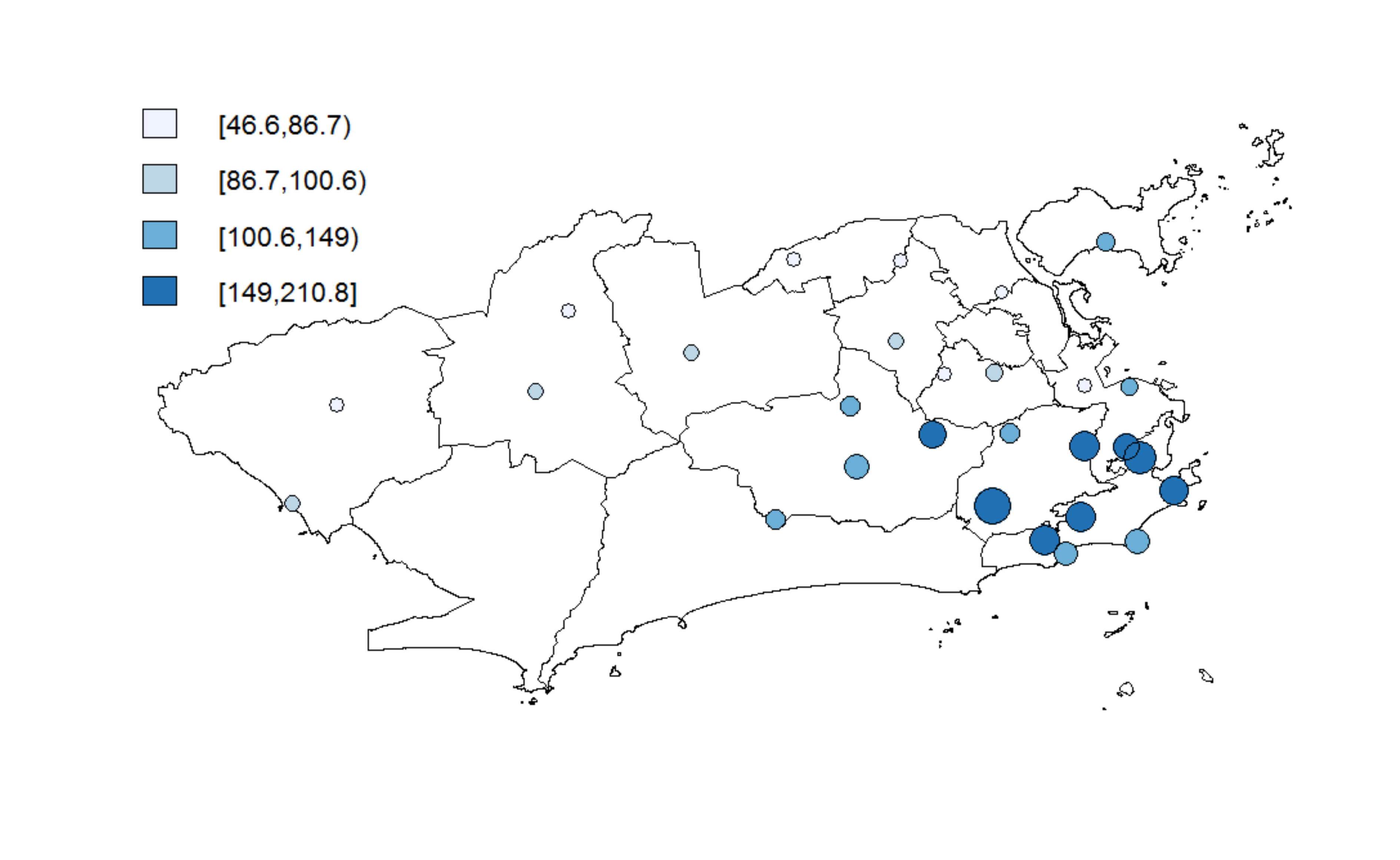}
\end{tabular}
\caption{Rainfall data: stations installed in Rio de Janeiro (the monitoring stations are separated according to the intensity of rainfall).}
\label{fig:fig6.1}
\end{center}
\end{figure}

Figure \ref{fig:fig6.1} presents the spatial arrangement of rainfall stations in Rio de Janeiro city. Notice that the spatial arrangement of the monitoring stations seems to indicate a higher concentration in places where precipitation levels are very large. It appears that the point pattern associated with the stations has been observed from an inhomogeneous process.\\

For statistical inference purposes, the spatial mean was adjusted considering latitude and longitude as covariates. The mixture models considered were the Gaussian (M1) and Gaussian-Log-Gaussian (M3) models presented in Section \ref{SecSimulate}. We investigated the goodness of fit for rainfall spatial modelling via cross-validation for spatially correlated data.\\

First, we arbitrarily choose $n_T=26$ and $n_V=5$, for the training and validation samples, respectively. In addition, we use $I=500$ split vectors and $H=3$ independent MCMC samples simulated from SIR estimator. We adopted the Mahalanobis distance as discrepancy measure between two random vectors of the same distribution, taking into account the covariance matrix.\\

The analysis of the posterior distribution of spatial mean shows significantly different estimates for both models. The spatial mean for M3 is significantly lower than the spatial mean estimated by M1. Actually, this makes sense since the process for the data is inhomogeneous. We evaluated both models with exponential covariance structures for spatial dependence. \\

The SIR estimator requires less computation time. Table \ref{tab7} shows the performance of both models using the Mahalanobis distance. The estimates obtained for M3 are smaller than M1 for both estimators. This is due to the fact that the Gaussian-Log-Gaussian process proposed by \cite{PSteel06} is able to capture heterogeneity in space through a mixing process used to increase the Gaussian process variability, although it does not take into account the monitoring stations arrangement dependence with the total rainfall.

\begin{table}[H]
\caption{Rainfall Data: cross-validation using Mahalanobis distance for $n_T=26$ and $n_V=5$.}
\label{tab7}
\begin{center}
\begin{tabular}{lcc}
\hline\noalign{\smallskip}
&  MC & SIR  \\ 
\noalign{\smallskip}\hline\noalign{\smallskip}
M1  & 6.983 ($1.9\times 10^{-6}$)  &  6.419 (0.0010)   \\
M3&    4.985 ($1.0\times 10^{-5}$) &  4.805 (0.0004)  \\
\noalign{\smallskip}\hline
\end{tabular}
\end{center}
\end{table}

In addition, we take into account spatial heterogeneity using stratified cross-validation. We divide the spatial region into only two strata, arbitrarily. In the first stratum, the monitoring stations are closed and there is a higher concentration of total rainfall data. The other stratum is defined by the remainder of the locations, i.e., more distant locations with lower values of total precipitation. This design is called design A1. Table \ref{tab8} presents the sample arrangement given by design A1 with the weights, $w$ respectively. The results for {design A1 can been see in Table \ref{tab9}. Among the strata, the best performance is achieved by M3.\\ 

An important issue in using cross-validation is the training dataset size. If we have an acceptable amount of training data, the model is sufficiently informed by the training set. Furthermore, this model is able to reproduce the observed data. We arbitrarily choose a small training sample, i.e., $n_T=10$ and $n_V=21$. Again, we use $I=500$ split vectors and $H=3$ independent MCMC samples simulated from SIR estimator. It is expected that using a reduced training sample size might cause some impact on the estimation of model parameters. Table \ref{tab10} displays the values of the Mahalanobis distance for the models considered without taking into account the heterogeneity of the data. The results of our analysis suggest it is best to use a relatively large training sample for making cross-validation under our approach.\\

\begin{table}[H]
\caption{Stratified design A1: training and validation samples. }
\label{tab8}
\begin{center}
\begin{tabular}{c  c c c }
\hline\noalign{\smallskip}
\textit{strata} & $n_t$ & $n_v$ &$w$\\
\noalign{\smallskip}\hline\noalign{\smallskip}
1  & 9 &  4 & 0.4\\
2  & 12 & 6 & 0.6 \\
\textit{total}  & 21 &  10 & 1.0 \\
\noalign{\smallskip}\hline
\end{tabular}
\end{center}
\end{table}

\begin{figure}[H]
\begin{center}
\begin{tabular}{l}
\includegraphics[width=0.85\textwidth]{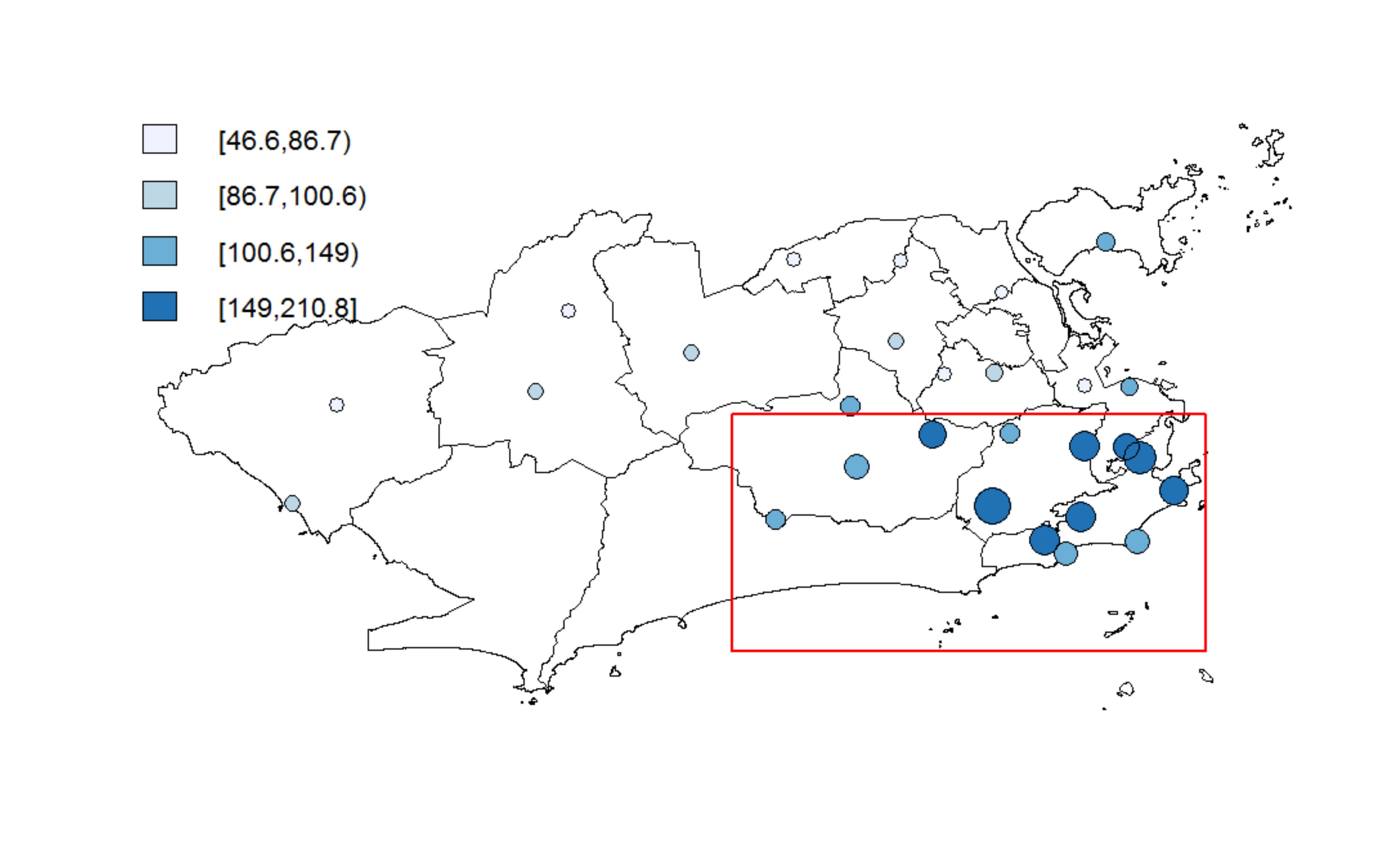}
\end{tabular}
\caption{Rainfall data: stations installed in Rio de Janeiro (the monitoring stations are separated according to the intensity of rainfall) for stratified design. Stratum 1 is represented by the red triangle and Stratum 2 by the remainder of the locations.}
\label{fig:fig6.2}
\end{center}
\end{figure}

\begin{table}[H]
\caption{Rainfall Data: cross-validation using Mahalanobis distance for $n_T=10$ and $n_V=21$.}
\label{tab10}
\begin{center}
\begin{tabular}{ccc}
\hline\noalign{\smallskip}
&  MC & SIR  \\ 
\noalign{\smallskip}\hline\noalign{\smallskip}
M1  & 15.257 ($7.4\times 10^{-6}$)  &14.699 (0.002)     \\
M3 &    7.829 ($5.6\times 10^{-7}$)   &   7.746 (0.003) \\
\noalign{\smallskip}\hline
\end{tabular}
\end{center}
\end{table}

\begin{table}[H]
\begin{center}
\caption{Rainfall Data:  stratified cross-validation using Mahalanobis distance for design A1.}
\label{tab9}
\begin{tabular}{l cccc}
\hline\noalign{\smallskip}
&  \multicolumn{2}{ c }{M1}  & \multicolumn{2}{ c }{M3} \\ 
$k$ &  MC & SIR & MC & SIR \\ 
\noalign{\smallskip}\hline\noalign{\smallskip}
1  &{4.485 ($1.3 \times 10^{-6}$)  } &   4.519 ($3.6 \times 10^{-4}$)  &  3.630 ($9.7 \times 10^{-7}$) &  3.520 ($1.4 \times 10^{-3}$)  \\ 
2  &{10.231 ($2.4 \times 10^{-6}$)} &  9.113 ($4.9 \times 10^{-4}$) &  2.563 ($3.1 \times 10^{-7}$)  &   2.666 ($5.6 \times 10^{-4}$) \\
\noalign{\smallskip}\hline\noalign{\smallskip}
$\hat{\Psi}^{st}$  &{7.933 ($1.9 \times 10^{-6}$)}  &   7.275 ($5.3 \times 10^{-5}$)   & 2.990 ($6.4 \times 10^{-7}$) &  3.007 ($1.3 \times 10^{-5}$)   \\
\noalign{\smallskip}\hline
\end{tabular}
\end{center}
\end{table}

\begin{table}[H]
\begin{center}
\caption{Rainfall Data:  stratified cross-validation using Mahalanobis distance for design A2.}
\label{tab12}
\begin{tabular}{lcccc}
\hline\noalign{\smallskip}
&  \multicolumn{2}{ c }{M1}  & \multicolumn{2}{ c }{M3} \\ 
$k$ &  MC & SIR & MC & SIR \\
\noalign{\smallskip}\hline\noalign{\smallskip}
1  &
 5.330 ($1.4\times 10^{-6}$)  &   5.317 ($3.0\times 10^{-3}$)   & 3.379 ($5.6\times 10^{-7}$)  &   4.485 ($2.1\times 10^{-4}$)  \\ 
2  &
12.583 ($3.7\times 10^{-6}$)  & 11.109 ($4.0\times 10^{-4}$) & 3.228 ($3.5\times 10^{-7}$) &   3.314 ($3.1\times 10^{-5}$) \\
\noalign{\smallskip}\hline
$\hat{\Psi}^{st}$  &
{9.681 ($2.5\times 10^{-6}$)}  &  8.792 ($5.1\times 10^{-4}$)    & 3.288 ($4.5\times 10^{-7}$) &  3.782 ($1.2\times 10^{-4}$)   \\
\noalign{\smallskip}\hline
\end{tabular}
\end{center}
\end{table}

Table \ref{tab11} presents the stratified design A2, considering a ``\textit{small}'' size for training sample. Observe that M1 produces high estimates in stratum 2 for both arrangements, as can be seem from Tables \ref{tab9} and \ref{tab12}. In fact, this might a result of the number of neighbours that are far apart.

\begin{table}[H]
\caption{Stratified design A2: training and validation samples. }
\label{tab11}
\begin{center}
\begin{tabular}{c c c c}
\hline\noalign{\smallskip}
 \textit{strata}  & $n_t$ & $n_v$ &$w$\\ 
\noalign{\smallskip}\hline\noalign{\smallskip}
1  & {7} &  6 & 0.4\\
2  & {9} & 9 & 0.6 \\
\textit{total}  & {16} &  15 & 1.0 \\
\noalign{\smallskip}\hline
\end{tabular}
\end{center}
\end{table}

It can be noted from Tables \ref{tab9} and \ref{tab12} that there is an increase in the accuracy of the SIR estimator by stratifying the spatial region. Furthermore, the stratification allows the identification of lack of fit for all models. \\

Note that M3 performs better for both designs indicating better predictive performance in all sub-regions considered in the two stratifications proposed in this application.

\section{Conclusions}

We consider Bayesian model comparison and criticism of geostatistical models. Cross-validation techniques are considered to evaluate the model predictive performances and we allow for uncertainty in the choice of validation sets through the prior distribution on the possible sets.\\

The results obtained in this study are useful for understanding the effect of cross-validation in spatial data analysis. This paper addresses important issues that have not been completely dealt with in the literature, such as the ad hoc choice of validation sets in spatial data analysis. \\

The proposed stratified reference distribution contributes to computing the stratified estimator and reduces the global variability in stratified cross-validation. The SIR estimator is a good approximation of the MC estimator and requires only a few MCMC runs for the parameter estimation step.\\

As pointed out by \cite{Cochran1999}, there are important issues related to the building of the strata, such as: as the potential variables used to determine them; the determination of their boundaries; and the number of strata. Moreover, the question of choosing the training sample size should also be considered and it is not trivial. We considered two different designs in the application to rainfall data to accommodate the possible effect of choosing a training set that is too small or too large. 

\section*{Acknowledgements}

This work was part of the Ph.D. research of V. G. R. Lobo under the supervision of T. C. O. Fonseca and F. A. S. Moura. V. G. R. Lobo benefited from a scholarship from {\em Conselho de Aperfei\c coamento de Pessoal de N\'ivel Superior} (CAPES), Brazil. T. C. O. Fonseca was partially supported by {\em Conselho Nacional de Desenvolvimento Cient\'ifico e Tecnol\'ogico } (CNPq). F. A. S. Moura was partially supported by CNPq.

\section*{\textbf{Appendix 1: Markov Chain Monte Carlo Sampler} \label{MCMC}}

\normalsize
The prior distributions considered for the parameters in Section \ref{sec2} and proposal densities used in the MCMC algorithm are detailed as follows.

\begin{enumerate}

\item  $\sigma^2 \sim GI(a,b)$, $a,b>0$. The proposal density in the MCMC sampler is: 
$$ ln(\sigma^2) \sim  Normal(ln(\sigma^{2(k-1)}),\sigma^2_{(\sigma^2)}). $$

\item $\mu \sim Normal_n( \mathbf{0}, \tau_{\mu}^2)$, $\tau_{\mu}^2>0$. The proposal density in the MCMC sampler is: $$ \mu \sim  Normal(\mu^{(k-1)},\sigma^2_{(\mu)}).$$

\item $\phi \sim Gama(1, c/med(u_s))$, with $c>0$ and $med(u_s)$ denoting the median distance in the observed data. The proposal density in the MCMC sampler is: $$ ln(\phi) \sim  Normal(ln(\phi^{(k-1)}),\sigma^2_{(\phi)}).$$

\item Jeffreys independent prior distribution \cite{Fonseca2008}: 
\small
$$  p(\nu) \propto \left(\frac{\nu}{\nu+3}\right)^{1/2}\left\{ \psi'\left(\frac{\nu}{2}\right) - \psi'\left(\frac{\nu+1}{2}\right) - \frac{2(\nu+3)}{\nu(\nu+1)^2}\right\}^{1/2},$$
\normalsize
with $\psi'(a)=\frac{d\left\{\psi(a)\right\}}{da}$ the trigamma function. In the context of regression models, this prior distribution guarantees that the posterior distribution for $\nu$ is proper. The proposal density in the MCMC sampler is: $$ ln(\nu) \sim  Normal(ln(\nu^{(k-1)}),\sigma^{2}_{(\nu)}) .$$
  
\end{enumerate}

\section*{\textbf{Appendix 2: Variance estimator} \label{APbronze}}

\normalsize

According to \cite{Robert}, the generic problem involves evaluating the integral

\begin{equation}
\label{eq:ap1}
E_f(h(X)) = \int_{\chi} h(x) f(x) dx,
\end{equation}

\noindent where $\chi$ denotes the set where the random variable $X$ takes its values, which is usually equal to the support of the density $f$. 

The principle of the Monte Carlo method for approximating equation \eqref{eq:ap1} is to generate a sample $X_1, \ldots, X_n$ from the density $f$ and proposed as an approximation to the empirical average

$$\bar{h}_n = \frac{1}{n} \sum_{j=1}^{n} h(x_j)$$
  
\noindent since $\bar{h}_n$ converges almost surely to $E_f(h(X))$ by the strong law of large numbers. 

When $h^2(X)$ has a finite expectation under $f$ the speed of convergence of $\bar{h}_n$ can be assessed, since the convergence takes place at a speed $O(\sqrt{n})$ and the asymptotic variance of the approximation is

\begin{equation}
\label{eq:ap2}
var(\bar{h}_n)= \frac{1}{n} \int_{\chi} \left[ h(x) - E_f(h(X)) \right]^2 f(x)dx, 
\end{equation}
  
  \noindent which can also be estimated from the sample $(X_1, \ldots, X_n)$ through

$$v_n= \frac{1}{n^2}\sum_{j=1}^{n} \left[ h(x_j) - \bar{h}_n \right]^2.$$
  
Analogously to equation \eqref{eq:ap2}, we can obtain the variance of the estimators $\hat{\Psi}_{mc}$ and $\hat{\Psi}_{sir}$. Notice that from the equation \eqref{eqS} we obtain,

\begin{equation}
\label{varS}
var(\hat{\Psi}_{mc}) = \frac{1}{I^2}\frac{1}{J^2}\sum_{i=1}^{I}\sum_{j=1}^{J} \left[r(\mathbf{s}_{y}^{(i)}, \mathbf{y}, \theta_{ij}, y_{ij}^{rep}) - \hat{\Psi}_{mc} \right]^2
\end{equation}

\noindent Thus, 
\small
\begin{equation}
\label{varB}
var(\hat{\Psi}_{sir}) = \frac{1}{H^2}  \frac{1}{I^2} \sum_{h=1}^{H} \sum_{i=1}^{I} \left[  \Psi_{hi} - \hat{\Psi}_{sir} \right]^2. 
\end{equation}

\normalsize
\noindent is the SIR estimator variance, obtained from equation \eqref{eqB}, where,

  $$ \Psi_{hi}= \frac{\sum_{j=1}^{J} r(\mathbf{s}_{y}^{(i)}, \mathbf{y}, \theta_{hj}, y_{hj}^{rep}) w_{hj}}{\sum_{j=1}^{J} w_{hj}} $$

\subsection*{\textbf{SIR estimator details}\label{APbronzeDetails}}

We draw a MCMC sample from $g(\theta)$, which is then reweighted using importance sampling to obtain $p(\theta \mid \mathbf{s}_{y})$. The same posterior sample serves for every split $\mathbf{s}_{y}$ considered, saving computational time. 

The equation weighting term $w_{hj}= \frac{p(\theta_{hj} \mid \mathbf{s}_{y}^{(i)},\mathbf{y})}{g(\theta_{hj})}$ can be obtained by applying the logarithm of the ratio as

\begin{eqnarray} \nonumber
log(w_{hj}) &=&log \left\{\frac{p(\theta_{hj} \mid \mathbf{s}_{y}^{(i)}, \mathbf{y})}{g(\theta_{hj})} \right\} = log \left\{ \frac{f(\mathbf{y}_{T[s]} \mid \theta_{hj})}{  f(\mathbf{y} \mid \theta_{hj})^{\alpha} } \right\} \\ 
&=& log \thinspace f(\mathbf{y}_{T[s]} \mid \theta_{hj}) - \alpha \thinspace  log f(\mathbf{y} \mid \theta_{hj})\\ \nonumber
\end{eqnarray}

\subsection*{\textbf{Stratified Variance $var(\hat{\Psi}_k)$}}

For the MC estimator, we have each $(\mathbf{s}_{y}^{(i)}, \mathbf{y}, \theta_{ij}, y_{ij}^{rep})$ distributed as the reference distribution. Then

$$ \hat{\Psi}_{{mc}_{k}} = \frac{1}{I_k}\sum_{i=1}^{I_k}\frac{1}{J}\sum_{j=1}^{J} r_{k}(\mathbf{s}_{y}^{(i)}, \mathbf{y}, \theta_{j}, y^{rep}_{ij})  $$
  
  \noindent is the MC estimator in each stratum. We can obtain the variance of the stratified MC estimator as

\begin{eqnarray} \nonumber \label{a2eq1}
var(\hat{\Psi}^{st}) &=& \frac{1}{n^2}\sum_{k=1}^{K} n_k(n_k - n_{V_{k}})\frac{s^2_k}{n_{V_{k}}} \\ \nonumber
&=& \sum_{k=1}^{K} \frac{w_k}{n}(n_k - n_{V_{k}})\frac{s^2_k}{n_{V_{k}}} \\ 
&=& \sum_{k=1}^{K} \frac{w_k}{n}(1 - f_{V_{k}})\frac{s^2_k}{n_{V_{k}}} \\ \nonumber
\end{eqnarray}

\noindent where

$$s^2_k=  \frac{1}{(n_{V_{k}}- 1)} \sum_{i=1}^{n_k} (r_{ki} - \hat{\Psi}_k)^2  $$
  
\noindent and $r_k$ represents any reference distribution. Note that equation \eqref{a2eq1} can be written as

\begin{eqnarray}\label{a2eq2} \nonumber
var(\hat{\Psi}^{st}) &=& var \left(\sum_{k=1}^{K} \hat{\Psi}^{st}_{k} \right)  \\ \nonumber
&=&  var \left(\sum_{k=1}^{K}  w_k \hat{\Psi}_{k} \right) \\ \nonumber
&=& \sum_{k=1}^{K} w_k^2 \thinspace var(\hat{\Psi}_k)  \\
\end{eqnarray}

Therefore, $var(\hat{\Psi}^{st}_{k}) = var(w_k \hat{\Psi}_{k}  )= w_k^2 \thinspace var( \hat{\Psi}_{k}  )$,$ \thinspace \forall \thinspace  k= 1, \ldots, K$. Analogously, we have a similar result for the SIR estimator variance.


\newcommand{\sortnoop}[1]{}

\end{document}